\documentclass[12pt,a4paper]{article}
\usepackage{defns}
\usepackage{ajr}

\newcommand{\bfa}[1]{\mbox{\boldmath $ #1 $}}
\newcommand{\bbu}{\bar{\bfa u}}
\newcommand{\re}{{\cal R}}
\newcommand{\we}{{\cal W}}
\newcommand{\bu}{{\bar u}}
\newcommand{\bv}{{\bar v}}
\newcommand{\bw}{{\bar\omega}}
\newcommand{\bd}{{\bar\delta}}
\newcommand{\gr}{{\cal G}}

\begin{document}

\title{The accurate and comprehensive model of thin fluid flows 
with inertia on curved substrates}

\author{A.~J. Roberts \and Zhenquan Li\thanks{Dept of Mathematics \&
Computing, University of Southern Queensland, Toowoomba, Queensland
4350, \textsc{Australia}; \protect\url{mailto:aroberts@usq.edu.au} and
\protect\url{mailto:zhen@usq.edu.au} respectively.} }

\maketitle
    
\begin{abstract}
Consider the 3D flow of a viscous Newtonian fluid upon a curved 2D
substrate when the fluid film is thin as occurs in many draining,
coating and biological flows.  We derive a comprehensive model of the
dynamics of the film, the model being expressed in terms of the film
thickness~$\eta$ and the average lateral velocity~$\bbu$.  Based upon
centre manifold theory, we are assured that the model accurately
includes the effects of the curvature of substrate, gravitational body
force, fluid inertia and dissipation.  The model may be used to resolve
wave-like phenomena in the dynamics of viscous fluid flows over
arbitrarily curved substrates such as cylinders, tubes and spheres.  We
briefly illustrate its use in simulating drop formation on cylindrical
fibres, wave transitions, Faraday waves, viscous hydraulic jumps, and
flow vortices in a compound channel.  These models are the most complete
models for thin film flow of a Newtonian fluid; many other thin film
models can be obtained by different truncations of the dynamical
equations given herein.
\end{abstract}

\tableofcontents

\section{Introduction}
\label{insu}

Mathematical models and numerical simulations for the flow of a thin
film of fluid have important applications in industrial and natural
processes \cite{Ruschak85, Roskes69, Schwartz95, Schwartz95b, Chang94,
Grotberg94, Moriarty91, Decre03}.  The dynamics of a thin fluid film
spreading or retracting from the surface of a supporting liquid or
solid substrate has long been active areas of research because of its
impact on many technological fields, for example, coating
flows~\cite{Ruschak85}: applications of coating flows range from a
single decorative layer on packaging, to multiple layer coatings on
photographic film; coated products include adhesive tape, surgical
dressings, magnetic and optical recording media, lithographic plates
paper and fabrics.  A wide variety of thin fluid film models have been
reviewed in detail by Oron \etal~\cite{Oron97}.  In this Introduction,
we summarise some of the results on mathematical models for three
dimensional thin fluid film flows on a solid curved substrate and
relate these to the new model derived herein.
   
In a three dimensional and very slow flow, a ``lubrication'' model for 
the evolution of the thickness~$\eta$ of a film on a general curved 
substrate was shown by Roy \etal~\cite{Roy96} to be
\begin{eqnarray}
\D{t}{\zeta}\approx-\third\we{\nabla}
     \cdot\left[\eta^2\zeta{\nabla }\tilde{\kappa}
      -\frac{1}{2}\eta^4(\kappa {\bfa I}-{\bfa K})
     \cdot{\nabla }\kappa\right]\,,
\label{roy}
\end{eqnarray}
where~$\zeta=\eta-\frac{1}{2}\kappa\eta^2+\frac{1}{3}k_1k_2\eta^3$ is
proportional to the amount of fluid locally ``above'' a small patch of
the substrate; $\tilde{\kappa}$ is the mean curvature of the free
surface of the film due to both substrate and fluid thickness
variations ($\tilde{\kappa}\approx\kappa+\nabla^2\eta$); ${\bfa K}$ is
the curvature tensor of the substrate; $k_1$, $k_2$ and
$\kappa=k_1+k_2$ are the principal curvatures and the mean curvature of
the substrate respectively (positive curvature is concave); $\we$ is a
Weber number characterising the strength of surface tension; and the
differential operator~$\nabla$ is defined in a coordinate system on the
curved substrate.  Based upon a systematic analysis of the continuity
and Navier-Stokes equations for a Newtonian fluid, this model accounts
for any general curvature of the substrate and that of the surface of
the film.  Decr\'e and Baret~\cite{Decre03} found good agreement
between a linearised version of lubrication models such as~(\ref{roy})
and experiments of flow over various shaped depressions in the
substrate.

\begin{figure}[tbp]
    \centering
    \begin{tabular}{cc}
        \includegraphics[width=0.43\textwidth]{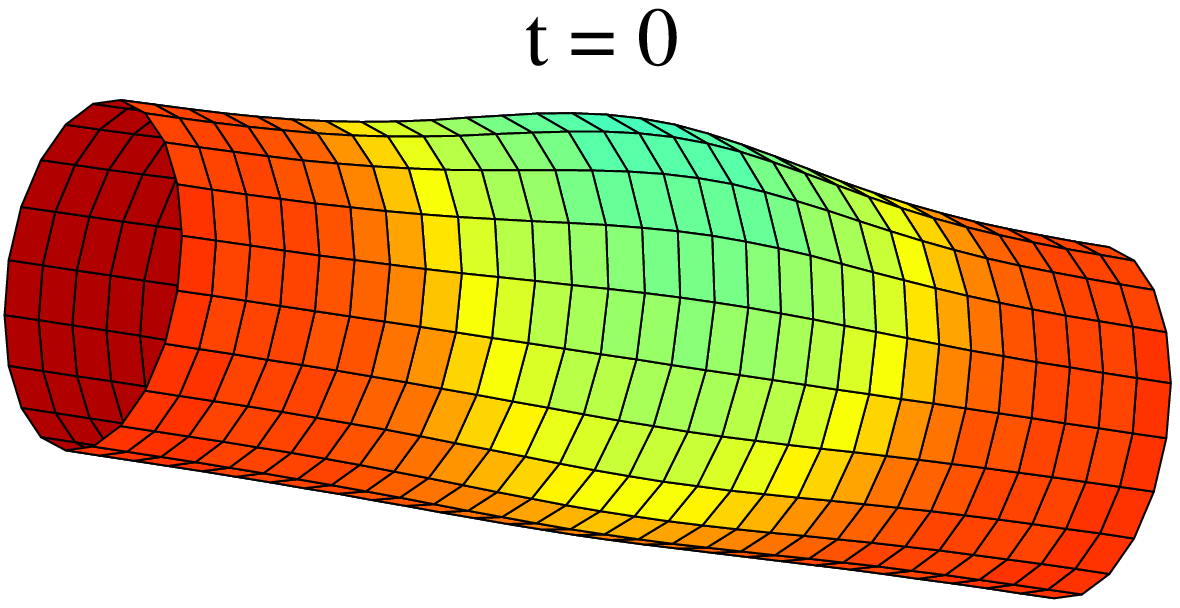} & 
        \includegraphics[width=0.43\textwidth]{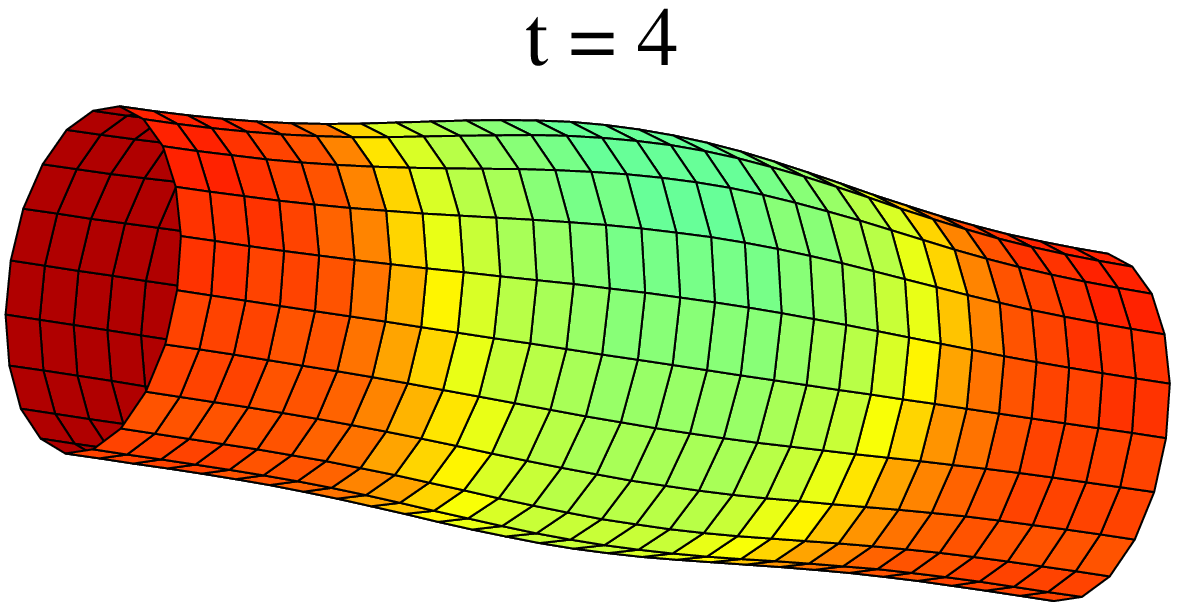}  \\
        \includegraphics[width=0.43\textwidth]{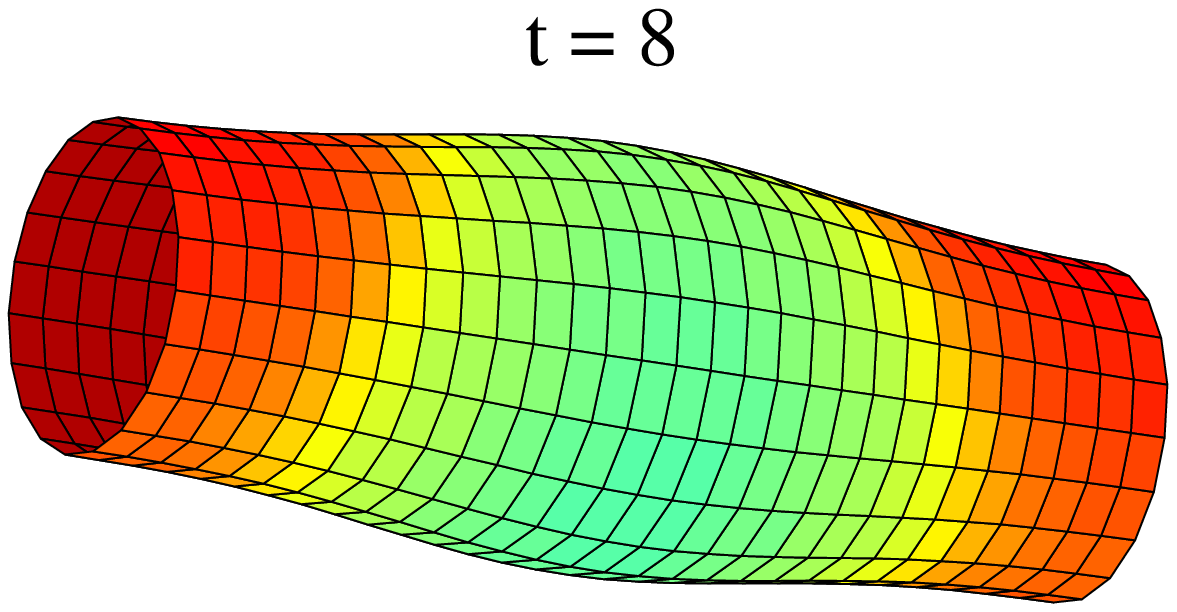} & 
        \includegraphics[width=0.43\textwidth]{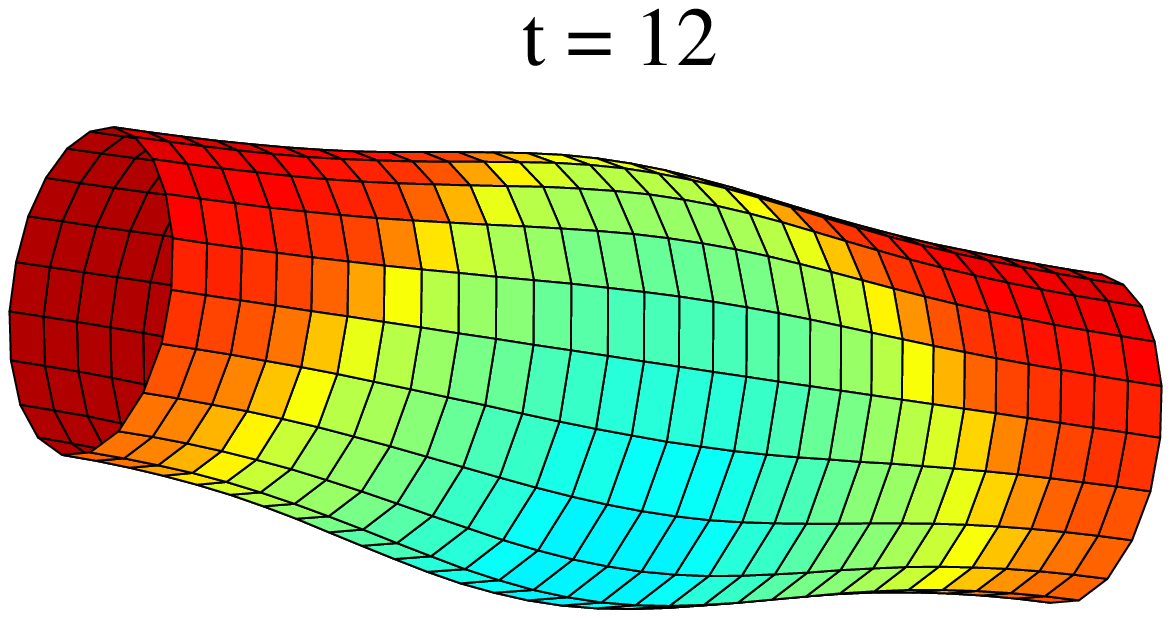}
    \end{tabular}
	\caption{around a nearly horizontal cylinder of radius $R=2$ (not
	shown) we start with a fluid layer of thickness~$\eta=1$ except for
	a small bump discernible off top dead centre of the cylinder.  With
	Reynolds number~$\re=10$ in a gravitational field with gravity
	number~$\gr=0.5$ the fluid bump first slides around the cylinder
	to the bottom by time~$t=12$\,.  The other sections of the fluid
	also slide down to the bottom of the cylinder, but not so fast.}
    \label{fig:cyl4}
\end{figure}

\begin{figure}[tbp]
    \centering
    \begin{tabular}{cc}
        \includegraphics[width=0.43\textwidth]{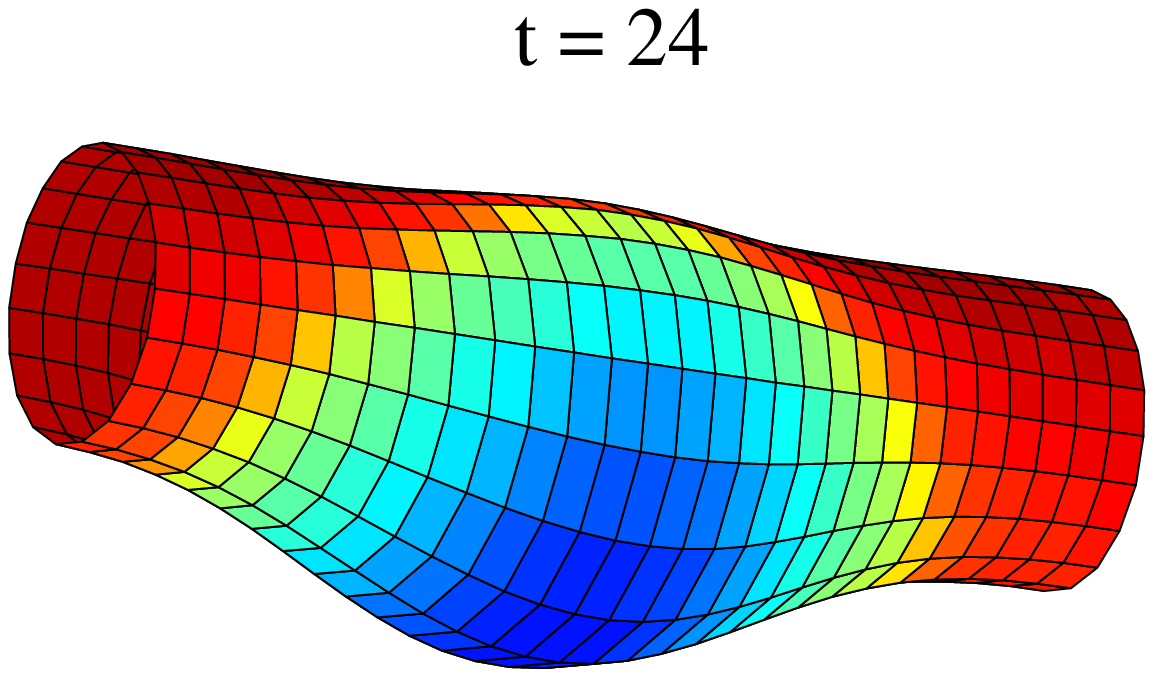} & 
        \includegraphics[width=0.43\textwidth]{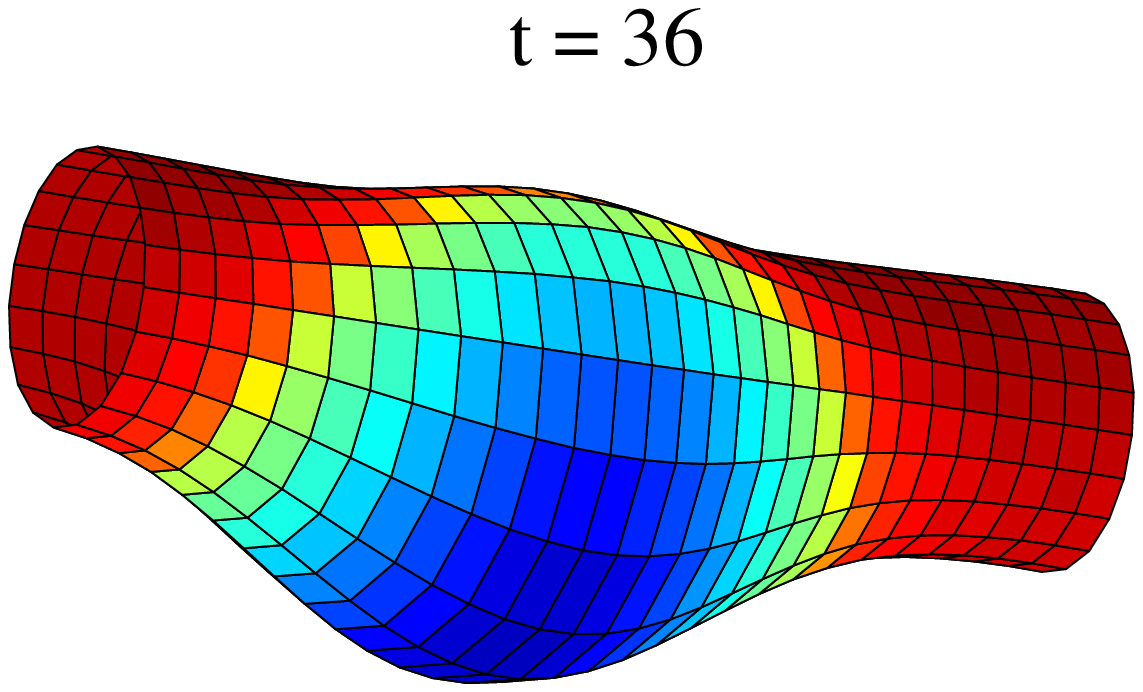}  \\
        \includegraphics[width=0.43\textwidth]{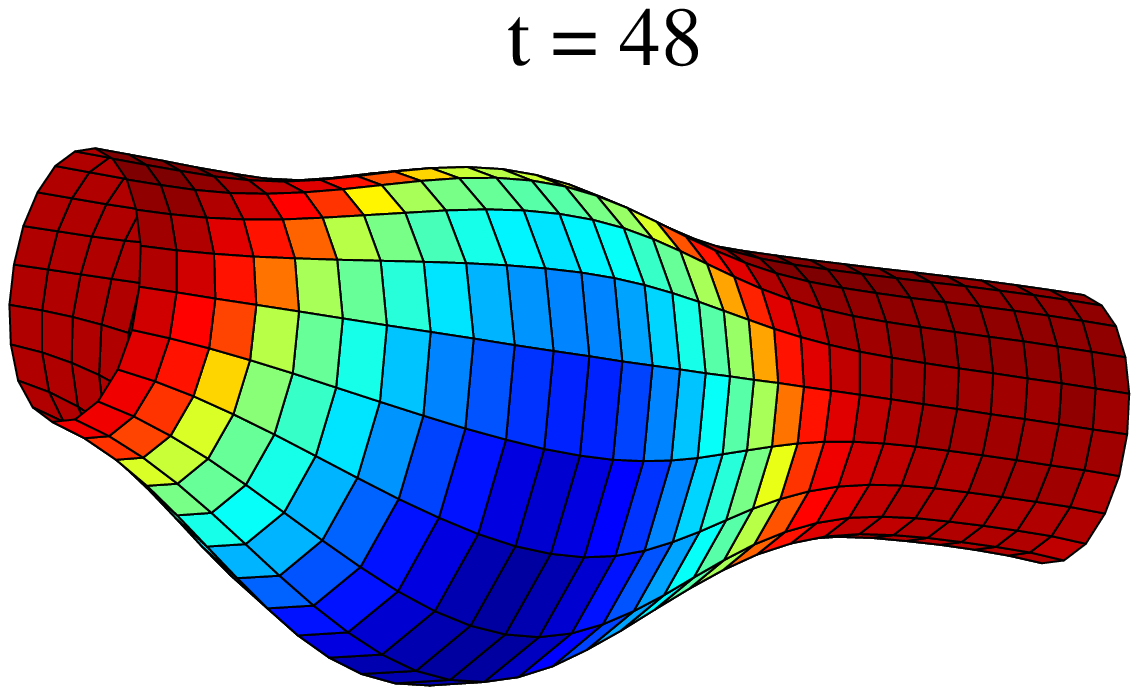} & 
        \includegraphics[width=0.43\textwidth]{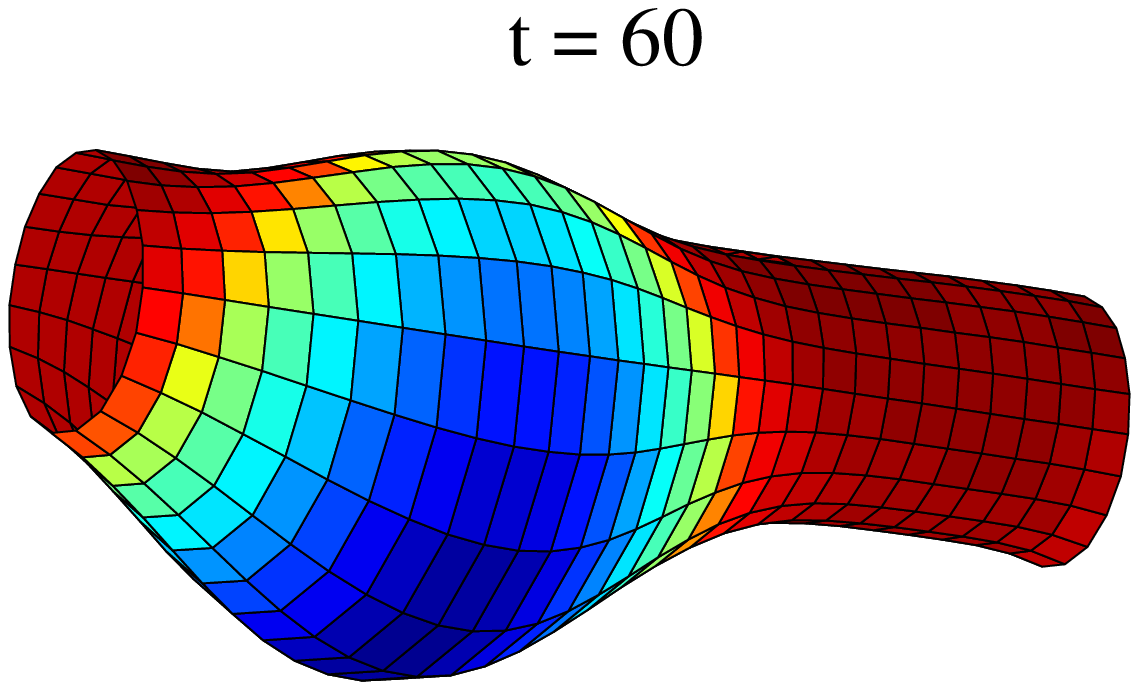}
    \end{tabular}
	\caption{around the nearly horizontal cylinder of radius $R=2$ (not
	shown but at angle $0.1$~radians to the horizontal) the fluid lump
	now, $t=24$\,, at the bottom of the cylinder slowly pulls in fluids
	from the two ends of the cylinder under surface tension of Weber
	number $\we=20$\,.  By times $t=48$ and~$60$ surface tension forms
	a large off-centre bead which slowly slides along the cylinder,
	surrounded by a thin layer, $\eta\approx0.1$\,, still covering the
	cylinder.}
    \label{fig:cyl8}
\end{figure}

In many applications the lubrication model~(\ref{roy}) of slow flow of
a thin fluid film has limited usefulness; instead a model expressed in
terms of both the fluid layer thickness and an overall lateral velocity or
momentum flux is needed to resolve faster wave-like dynamics in falling
films~\cite[p110]{Chang94}, wave transitions~\cite{Chang02}, higher
Reynolds number flows~\cite[Eqn.(19)]{Prokopiou91b}, in rising film
flow and a slot coater~\cite[Eqn.(37)]{Kheshgi89}, and in general
flows~\cite{Haragus95, Roberts96b}.  Roberts~\cite{Roberts96b} derived
such a model for two dimensional flow, approximately
\begin{eqnarray}
\D{t}{\eta}&\approx&-\D x{(\eta\bar{u})}\,, 
\label{Roberts1}
\\
\re\D{t}{\bar{u}}
         &\approx&-\left[\frac{\pi^2}{4}\frac{\bar{u}}{\eta^2}
	 +3{\kappa}\frac{\bar{u}}{\eta}\right] 
+\frac{\pi^2}{12}\left(\we\D{x}{\tilde{\kappa}}+\gr g_s\right)\,,
\label{Roberts2}
\end{eqnarray}
where $\re$ is a Reynolds number of the flow, $\bar{u}$~is the lateral
velocity averaged over the fluid thickness, and $\gr g_s$~is the
lateral component of gravity.  Here we greatly extend the
model~(\ref{Roberts1}--\ref{Roberts2}) by deriving in
\S\ref{ldm3d} the approximate model for the flow of a three
dimensional thin liquid layer of an incompressible, Newtonian fluid
over an arbitrary solid, stationary and curved substrate, such as the 
flow about a cylinder shown in Figures~\ref{fig:cyl4}
and~\ref{fig:cyl8}.
The derived accurate model~(\ref{ldm3d1}--\ref{ldm3d2}) for the film
thickness~$\eta$ and a weighted average lateral velocity~$\bar{\bfa
u}$, see~(\ref{eq:baru}), is to low order and analogous
to~(\ref{Roberts1}--\ref{Roberts2})\footnote{The approximate equality
in the conservation of mass equation~(\ref{my1}) becomes exact equality
when $\zeta$ replaces $\eta$~on the left-hand side.  The higher order
analysis leading to~(\ref{ldm3d1}) does this automatically.}
\begin{eqnarray}
\D{t}{\eta}&\approx&
-{\nabla}\cdot(\eta\bbu)\,, 
\label{my1}
\\
\re\D{t}{\bbu}
         &\approx&-\left[\frac{\pi^2}{4}\frac{\bbu}{\eta^2}
         +\left(2{\bfa K}+{\kappa}{\bfa I}\right)\cdot 
   \frac{\bbu}{\eta}\right]+\frac{\pi^2}{12}\left(\we{\nabla}
   \tilde{\kappa}+\gr{\bfa g}_s\right)\,, 
\label{my2}
\end{eqnarray}
where $\gr{\bfa g}_s$ is the component of gravity tangential to the
substrate.  The conservation of fluid equation~(\ref{my1}) is the
natural generalisation of equation~(\ref{Roberts1}) to three
dimensional flow.  The momentum equation~(\ref{my2}) similarly
generalises~(\ref{Roberts2}) to three dimensional flow through a
nontrivial effect of curvature upon drag.  The important feature of
this model, as in~(\ref{Roberts1}--\ref{Roberts2}), is the
incorporation of the dynamics of the inertia of the fluid, represented
here by the leading order term $\re{\partial \bbu}/{\partial t}$\,,
which enables the model to resolve wave-like behaviour.  In contrast,
the lubrication model of thin films~(\ref{roy}) only encompasses a much
more restricted range of dynamics.  We base the derivation of
model~(\ref{my1}--\ref{my2}) upon a centre manifold approach
established in \S\ref{cmam}.  The approach is founded on viscosity
damping all the lateral shear modes of the thin fluid film except the
shear mode of slowest decay.  Then {all} the physical interactions
between spatial varying quantities, substrate curvature, surface
tension and gravitational forcing are systematically incorporated into
the modelling because the centre manifold is made up of the slowly
evolving solutions of the Navier-Stokes and continuity equations; for
example, all these physical effects take part in the particular
simulation shown in Figures~\ref{fig:cyl4} and~\ref{fig:cyl8}.  For
example, the $\pi^2/12$ coefficient in the models such
as~(\ref{Roberts2},\ref{my2}) is not~$1$: the coefficient of these
terms would be~$1$ in modelling based upon the heuristic of cross
sectional averaging; but $\pi^2/12=0.8224$ is correct because it must
be $\third$ of the viscous decay rate~$\pi^2/4$ in order to match the
leading $\third$~coefficient in the lubrication model~(\ref{roy}). In
this approach the model is based upon actual solutions of the
Navier-Stokes equations and so gets all coefficients correct to a
controlable order of accuracy.

Before we undertake the construction we introduce in \S\ref{ortcur} an
orthogonal curvilinear coordinate system fitted to the substrate.  The
analysis then starts from the incompressible Navier-Stokes equations
and boundary conditions recorded in \S\ref{embc} for this special
coordinate system.  Thus we derive the model for a completely general
curving substrate.

The derived model~(\ref{ldm3d1}--\ref{ldm3d2}) reduces to a model 
for three dimensional fluid flow on flat substrates upon setting the 
principle curvatures~$k_1$ and~$k_2$ equal to zero.  For example, the 
low order model~(\ref{my2}) becomes simply
\begin{equation}
\re\D{t}{\bbu}
         \approx -\frac{\pi^2}{4}\frac{\bbu}{\eta^2}
         +\frac{\pi^2}{12}\left(\we{\nabla}^3\eta+\gr{\bfa g}_s\right)\,.    
    \label{eq:lowflat}
\end{equation}
The higher order accurate version of this model, recorded in 
\S\ref{Sflat} as~(\ref{eq:flath}--\ref{eq:flatv}), extends to 
three dimensional fluid flows the models for two-dimensional fluids on 
flat substrates that were derived in~\cite{Roberts96b, Haragus95}.
In \S\ref{Sflat} we report on the linearised dynamics, $\eta=1+h$ 
where both~$h$ and~$\bbu$ are small.
One result is that
\begin{equation}
    \re\bw_t=-\frac{\pi^2}{4}\bw+\nabla^2\bw\,,
    \label{eq:vvort}
\end{equation}
where $\bw=\bv_x-\bu_y$ is a measure of the mean vorticity normal to 
the substrate in the flow of the film.
Thus mean normal vorticity just dissipates due to drag and diffusion.
However, letting $\bd=\bu_x+\bv_y$\,, which measures the mean divergence 
of the flow of the film and hence indicates whether the film is 
thinning or thickening, we find
\begin{eqnarray}
    h_t & = & -\bd\,,
    \label{eq:mcont}  \\
    \re\bd_t & = & -\frac{\pi^2}{4}\bd
    +\frac{\pi^2}{12}\left[ \gr g_n\nabla^2h+\we\nabla^4h \right]
    +(1+\varpi)\nabla^2\bd\,,
    \label{eq:mdiv}
\end{eqnarray}
where $\varpi=3.0930$\,.  Observe that this divergence diffuses with a
larger coefficient, namely $1+\varpi$\,, than that of pure molecular
diffusion; this effect is analogous to the enhanced Trouton viscosity
of deforming viscous sheets~\cite[p143, e.g.]{Ribe01}.  The enhanced
viscous dissipation is due to interactions with the shear flow similar
to those giving rise to enhanced dispersion of a passive tracer in
pipes~\cite[e.g.]{Mercer94a}.  From~(\ref{eq:mdiv}) see that the
divergence of the film's velocity is simply driven by gravity and surface
tension acting on variations of the film's thickness and is dissipated
by substrate drag and the enhanced lateral diffusion.  Nonlinearities
and substrate curvature modify this simple picture of the dynamics.

Circular cylinders are a specific substrate of wide interest.  For
example, Jensen~\cite{Jensen97} studied the effects of surface tension
on a thin liquid layer lining the interior of a cylindrical tube and
derived a corresponding evolution equation.  Whereas Atherton \&
Homsy~\cite{Atherton76}, Kalliadasis \& Chang~\cite{Kalliadasis94} and
Kliakhandler \etal~\cite{Kliakhandler01} considered coating films flow
down vertical fibres and gave nonlinear lubrication models.  Thus in
\S\ref{Scyl} we also record the accurate model for flow both inside and
outside a cylinder as used in the simulations for
Figures~\ref{fig:cyl4} and~\ref{fig:cyl8}.  Axisymmetric flows are
often of interest in coating flows.  Using~$s$ as the axial coordinate,
to low order a model for axisymmetric film flow along a cylinder of
radius~$R$ is
\begin{eqnarray}
     &&   \D t{}\left(\eta\pm\frac{\eta^2}{2R}\right) = 
	-\D s{(\eta\bu )}\,,
    \label{eq:axiloh}  \\&&
    \re\D t{\bu } \approx 
    -\frac{\pi^2}{4}\frac{\bu }{\eta^2} 
    \pm\frac{\bu }{R\eta} -0.6487\frac{\bu }{R^2}
    \nonumber\\&&\quad{}
    +\frac{\pi^2}{12}\we\left(\frac{1}{R^2}\D s\eta +\DDD s\eta \right)
    +\gr\left(\frac{\pi^2}{12} g_s
    \pm 0.4891\frac{g_s\eta}{R}\right)\,,
    \label{eq:axilou}
\end{eqnarray}
where the upper/lower sign corresponds to flow on the outside/inside
surfaces of the cylinder.  Observe that the curvature of the substrate:
modifies the expression of conservation of mass; drives a beading
effect through $\eta_s/R^2$; and modifies the drag terms.  These are
just some special cases of those recorded in \S\ref{Sspec}.

The centre manifold approach we use to derive low-dimensional dynamical
models such as~(\ref{my1}--\ref{my2}) has the advantage that its simple
geometric picture leads to a complete low-dimensional
model~\cite[e.g.]{Mercer94a}.  Algebraic techniques have been developed
for the derivation of a low-dimensional model~\cite{Roberts96a}, the
correct modelling of initial conditions~\cite{Cox93b, Roberts89,
Suslov98b} and boundary conditions~\cite{Roberts92}.  However, we limit
our attention to deriving the basic differential equations of the
dynamical flow.  Other aspects of modelling remain for further study.
But further, at the end of \S\ref{ldm3d} we investigate high order
refinements of the basic linearised surface tension driven dynamics
of~(\ref{eq:vvort}--\ref{eq:mdiv}) and determine that the
model~(\ref{ldm3d1}--\ref{ldm3d2}) derived here requires that spatial
gradients are significantly less than the limit $|\grad\eta|<1.9$, for
example, see~(\ref{eq:logder}).  This quantitative indication of the
extent of the model's spatial resolution is better than
that for lubrication models such as~(\ref{roy}) which require the
surface slope to be significantly less than~$0.74$ instead.  Such
quantitative estimates of the range of applicability are found through
the systematic nature of the centre manifold approach to modelling.

\section{The orthogonal curvilinear coordinate system} 
\label{ortcur}
In this section, we describe the general differential geometry 
necessary to consider flows in general non-Cartesian geometries.
First, we introduce the geometry on the substrate, then extend it out 
into space and establish the orthogonal curvilinear coordinate system
used to describe the fluid flow.

Let $\cal S$ denote the solid substrate.
When $\cal S$ has no umbilical point, that is, there is no point on $\cal 
S$ at which the two principal curvatures coincide, then there are 
exactly two mutually orthogonal principal directions in the tangent 
plane at every point in ${\cal S}$~\cite[Theorem 10-3]{Gugg63}.
Let ${\bfa e}_1$ and ${\bfa e}_2$ be the unit vectors in these 
principal directions, and ${\bfa e}_3$ the unit normal to the 
substrate to the side of the fluid so that ${\bfa e}_1$, ${\bfa e}_2$ 
and ${\bfa e}_3$ form a right-handed curvilinear orthonormal set of 
unit vectors.
Such a coordinate system is called a Darboux frame~\cite{Gugg63}.
Let~$x_1$ and $x_2$ be two parameters such that the unit tangent vector 
of a parameter curve $x_2=$constant is ${\bfa e}_1$, the unit tangent 
vector of a parameter curve $x_1$=constant is ${\bfa e}_2$, and let 
$y$ measure the normal distance from the substrate.  Then on the 
substrate, points ${\bfa P}\in\cal S$\,,
\begin{equation}
{\bfa e}_i=\frac{1}{m_i}\D{x_i}{{\bfa P}}\,,
\label{units}
\end{equation}
with substrate scale factor 
\begin{equation}
m_i=\left|\D{x_i}{{\bfa P}}\right|\,.
\label{scalef}
\end{equation}
Further, we deduce how the normal varies along the substrate:
\begin{equation}
\D{x_i}{{\bfa e}_3}=-m_ik_i{\bfa e}_i\,.
\label{eqpav}
\end{equation}
Note that the unit vectors~${\bfa e}_i$ are independent of~$y$.  At 
any point in the fluid, written as
\begin{displaymath}
    {\bfa X}(x_1,x_2,y)={\bfa P}(x_1,x_2)+y{\bfa e}_3(x_1,x_2)\,, 
\end{displaymath}
the scale factors of the spatial coordinate system are, since positive 
curvature corresponds to a concave coordinate curve,
\begin{displaymath}
h_i=\left|\D{x_i}{{\bfa X}}\right|
=m_i(1-k_iy)\,,\quad 
h_3\left|\D{y}{{\bfa X}}\right|
=1\,. 
\end{displaymath}
The spatial derivatives of the curvilinear unit vectors 
are~\cite[p598]{Batchelor79} 
\begin{eqnarray*}
    && \D{ x_i}{{\bfa 
e}_i}=-\frac{h_{i,i'}}{h_{i'}}{\bfa e}_{i'}+m_ik_i{\bfa 
e}_3\,,\quad\quad \D{y}{{\bfa e}_i}=\D y{{\bfa e}_3}=\bfa 0\,,
\\&& 
\D{x_i}{{\bfa e}_3}=-m_ik_i{\bfa 
e}_i\,,\,\quad\quad\quad\quad\quad\quad \D{x_{i'}}{{\bfa 
e}_i}=\frac{h_{i',i}}{h_i}{\bfa e}_{i'}\,.
\end{eqnarray*}
where $i'=3-i$ is the complementary index of~$i$ and~$h_{i,j}$ denotes 
${\partial h_i}/{\partial x_j}$\,.

A fundamental quantity for the problem is the free-surface mean 
curvature~$\tilde{\kappa}$ which is involved in the effects of surface 
tension through the energy stored in the free-surface.
As derived by Roy \etal~\cite[eqn(37)]{Roy96}, the mean 
curvature of the free-surface 
\begin{eqnarray*}
\tilde{\kappa}&=&\frac{1}{\tilde{h}_1\tilde{h}_2}\left[\D{x_1}{}
\left(\frac{\tilde{h}_2^2\eta_{x_1}}{\cal A}\right)+\D{x_2}{}
\left(\frac{\tilde{h}_1^2\eta_{x_2}}{\cal A}\right)\right] 
\nonumber\\
& &{}+\frac{1}{\cal A}\left[\left(\tilde{h}_1^2
+\eta_{x_1}^2\right)\frac{m_2k_2}{\tilde{h}_1}
+\left(\tilde{h}_2^2+\eta_{x_2}^2\right)\frac{m_1k_1}{\tilde{h}_2}\right]\,, 
\nonumber
\end{eqnarray*}
where \mbox{$\tilde{h}_i=1-k_i \eta$} are the metric coefficients
evaluated on the free surface (indicated by the tilde), and where 
\begin{center}
${\cal A}=\sqrt{\tilde{h}_1^2\tilde{h}_2^2+\tilde{h}_2^2\eta_{x_1}^2
+\tilde{h}_1^2\eta_{x_2}^2}$\,, 
\end{center}
is proportional to the free-surface area above a patch~$dx_1\times 
dx_2$ of the substrate.

In the analysis we assume the film of fluid is thin.  Conversely,
viewing this on the scale of its thickness, the viscous fluid is of
large horizontal extent on a slowly curving substrate.  Thus we treat
as small the spatial derivatives of the fluid flow and the curvatures
of the substrate; Decr\'e \& Baret~\cite[p162]{Decre03} report
experiments showing that apparently rapid changes in the substrate may
be treated as slow variations in the mathematical model.  Then an
approximation to the curvature of the free surface is
\begin{equation}
\tilde{\kappa}=\nabla^2\eta+\frac{k_1}{1-k_1\eta}+\frac{k_2}{1-k_2\eta}
+\Ord{\kappa^3+\nabla^3\eta}\,, 
\label{eq9.1}
\end{equation}
where the Laplacian is evaluated in the substrate coordinate system 
as 
\begin{eqnarray}
\nabla^2\eta=\frac{1}{{m}_1{m}_2}\left[\D{x_1}{}
              \left(\frac{{m}_2}{{m}_1}\D{
x_1}{\eta}\right)+
              \D{x_2}{}\left(\frac{{m}_1}{{m}_2}
              \D{x_2}{\eta}\right)\right]\,.
\label{surfna}
\end{eqnarray}
For later use, also observe that on the free-surface the unit tangent 
vectors~$\tilde{\bfa t}_i$ and unit normal vector~$\tilde{\bfa n}$ are
\begin{equation}
\tilde{\bfa t}_i=(\tilde{h}_i{\bfa e}_1+\eta_{x_i}{\bfa 
e}_3)/\sqrt{{\tilde{h}_i}^2+\eta_{x_i}^2}\,, 
\label{tvect}
\end{equation}
\begin{equation}
\tilde{\bfa n}=(-\tilde{h}_2\eta_{x_1}{\bfa 
e}_1-\tilde{h}_1\eta_{x_2}{\bfa e}_2
+\tilde{h}_1\tilde{h}_2{\bfa e}_3)/{\cal A}\,. 
\label{nvect}
\end{equation}
We describe the dynamics of the fluid using these formula in a
coordinate system determined by the substrate upon which the film
flows.

\section{Equations of motion and boundary conditions}
\label{embc}

Having developed the intrinsic geometry of general three dimensional 
surfaces, we proceed to record the Navier-Stokes equations and 
boundary conditions for a Newtonian fluid in this curvilinear 
coordinate system.

Consider the Navier-Stokes equations for an incompressible fluid flow 
moving with velocity field~${\bfa u}$ and with pressure field~$p$.
The flow dynamics are driven by pressure gradients along the substrate 
which are caused by both surface tension forces, coefficient~$\sigma$, 
the forces varying due to variations of the curvature of the free 
surface of the fluid, and a gravitational acceleration, ${\bfa g}$, of 
magnitude~$g$ in the direction of the unit vector~$\hat{\bfa g}$.
Then the continuity and Navier-Stokes equations are
\begin{equation}
\nabla\cdot{\bfa u} = 0\,,
\end{equation}
\begin{equation}
\D{t}{{\bfa u}}+{\bfa u}\cdot\nabla{\bfa u}
 =  -\frac{1}{\rho}\nabla p+\frac{\mu}{\rho}\nabla^2{\bfa u}+{\bfa 
g}\,.
\end{equation}
We adopt a non-dimensionalisation based upon the characteristic
thickness of the film~$H$, and some characteristic velocity~$U$: for a
specific example, in a regime where surface tension drives a flow
against viscous drag the characteristic velocity is~$U=\sigma/\mu$ and
the Weber number~$\we=\sigma/(U\mu)=1$\,.  Reverting to the general
case, the reference length is~$H$, the reference time~$H/U$, and the
reference pressure~$\mu U/H$.  Then the non-di\-men\-sion\-al fluid
equations are
\begin{equation}
\nabla\cdot{\bfa u} = 0\,,
\end{equation}
\begin{equation}
\re\left[\D{t}{{\bfa u}}+{\bfa 
u}\cdot\nabla{\bfa u}\right]
 =  -\nabla p+\nabla^2{\bfa u}+\gr \hat{\bfa g}\,, 
 \label{eq:nonns}
\end{equation}
where~$\re=UH\rho/\mu$ is a Reynolds number characterising the
importance of the inertial terms compared to viscous dissipation, and
$\gr =g \rho H^2/(\mu U)$ is a gravity number analogously measuring the
importance of the gravitational body force compared to viscous
dissipation; the gravity number~$\gr=\re/{\cal F}$ for Froude number
${\cal F}=U^2/gH$ so that when we choose the reference velocity to be
the shallow water wave speed $U=\sqrt{gH}$ then ${\cal F}=1$ and the
gravity number $\gr=\re$\,.

In the curvilinear coordinate system defined in \S\ref{ortcur} 
the non-di\-men\-sion\-al continuity and Navier-Stokes equations for the 
velocity field~${\bfa u}= u_1 {\bfa e}_1+ u_2 {\bfa e}_2+ v 
{\bfa e}_3$ are as follows (adapted from~\cite[p599]{Batchelor79}):
\begin{eqnarray} 
    &&
\D{x_1}{}(h_2 u_1)+\D{x_2}{}(h_1u_2)+
\D{y}{}(h_1h_2 v)=0\,, 
\label{curequ1}\\&&
\re\left\{\D{t}{{\bfa u}}
+{\bfa e}_1\left[{\bfa u}\cdot\nabla 
u_1+\frac{u_2}{h_1h_2}\left(u_1\D{
x_2}{h_1}-u_2\D{
x_1}{h_2}\right)-m_1k_1\frac{vu_1}{h_1}\right]\right.
\nonumber\\&&\quad{}
+{\bfa e}_2\left[{\bfa u}\cdot\nabla 
u_2+\frac{u_1}{h_1h_2}\left(u_2\D{x_1}{h_2}-u_1\D{
x_2}{h_1}\right)-m_2k_2\frac{vu_2}{h_2}\right] 
\label{curequ2}\\&&\quad{}
+\left.{\bfa e}_3\left[{\bfa u}\cdot\nabla 
v+m_1k_1\frac{{u_1}^2}{h_1}+m_2k_2\frac{{u_2}^2}{h_2}\right]\right\}
=-\nabla p-\nabla\times{\bfa \omega} 
+\gr \hat{\bfa g}\,,
\nonumber
\end{eqnarray}
where~${\bfa \omega}$ is the vorticity of the fluid given by the curl
\begin{eqnarray*}
{\bfa \omega}=\nabla\times{\bfa u}&=&\frac{{\bfa 
e}_1}{h_2}\left[\D{x_2}{v}-\D y{(h_2u_2)}\right]
+\frac{{\bfa e}_2}{h_1}\left[\D{
y}{(h_1u_1)}-\D{x_1}{v}\right]
\\&&{}
+\frac{{\bfa e}_3}{h_1h_2}\left[\D{
x_1}{(h_2u_2)}-\D{x_2}{(h_1u_1)}\right]\,,
\end{eqnarray*}
and where
\begin{eqnarray*}
{\bfa u}\cdot\nabla=\frac{u_1}{h_1}\D{
x_1}{} +\frac{u_2}{h_2}\D{x_2}{}
+v\D{y}{}\,.
\end{eqnarray*}
These partial differential equations are solved with the following 
boundary conditions.
\begin{enumerate}
\item
The fluid does not slip along the stationary substrate, that is
\begin{equation}
{\bfa u}={\bfa 0}\quad\mbox{on $y=0$}\,. 
\label{bcon1}
\end{equation}
\item
The fluid satisfies the free surface kinematic boundary condition
\begin{equation}
\D{
t}{\eta}=v-\frac{u_1}{\tilde{h}_1}\D{x_1}{\eta}
-\frac{u_2}{\tilde{h}_2}\D{
x_2}{\eta}\quad\mbox{on $y=\eta$}\,.
\label{bcon2}
\end{equation}
\item
The normal surface stress at the free surface is caused by surface
tension, in non-di\-men\-sion\-al form
\begin{equation}
\tilde{\bfa n}\cdot\tilde{\bfa {\tau}}\cdot\tilde{\bfa 
n}=\tilde{p}+\we\tilde{\kappa}\,,
\label{bcon3}
\end{equation}
where $\tilde{\bfa {\tau}}$ is the deviatoric stress tensor on free 
surface, $\tilde{p}$ is the fluid pressure at the surface relative to 
the assumed zero pressure of the negligible medium above the fluid, 
and~$\tilde{\bfa n}$ is the unit normal to the free surface.
\item
The free surface has zero tangential stress
\begin{equation}
\tilde{\bfa t} \cdot\tilde{\bfa {\tau}}\cdot\tilde{\bfa n}=0\,, 
\label{bcon4}
\end{equation}
where $\tilde{\bfa t}$ is any tangent vector to the free surface.  We 
use the two linearly independent tangent vectors in~(\ref{tvect}) to 
ensure the boundary condition is satisfied for all tangent vectors.
\end{enumerate}
In this curvilinear coordinate system the components of the 
nondimensional deviatoric stress tensor~${\bfa \tau}$ 
are~\cite[p599]{Batchelor79}
\begin{eqnarray}
\tau_{ii} & = &2\left(\frac{1}{h_i}\D{x_i}{u_i}+
\frac{h_{i,i'}}{h_ih_{i'}}u_{i'}-\frac{m_ik_i}{h_i}v\right)\,,\nonumber 
\\
\tau_{12} & = &\frac{1}{h_2}\D{
x_2}{u_1}+\frac{1}{h_1}\D{
x_1}{u_2}-\frac{h_{1,2}}{h_1h_2}u_1-
\frac{h_{2,1}}{h_1h_2}u_2\,, \\
\tau_{i3} & = &\frac{1}{h_i}\D{
x_i}{v}+\D{
y}{u_i}+\frac{m_ik_i}{h_i}u_i\,,\nonumber\\
\tau_{33} & = & 2\D{y}{v}\,.\nonumber 
\end{eqnarray}

\section{The centre manifold analysis of the dynamics}
\label{cmam}
We adapt the governing fluid equations~(\ref{curequ1}--\ref{curequ2})
and the four boundary conditions~(\ref{bcon1}--\ref{bcon4}) to a form
suitable for the application of centre manifold theory and techniques
to generate a low-dimensional dynamical model with firm theoretical
support.

Three mathematical tricks place the equations within the centre
manifold framework; these tricks fit the parameter regime of
viscous flow varying relatively slowly over the substrate.  
\begin{enumerate}
	\item First we introduce the small parameter~$\epsilon$ to
	characterise both the slow gradients along the substrate,
	${\partial}/{\partial x_i}$\,, and the small curvatures of the
	substrate (as curvatures are the partial derivatives of unit normal
	with respect to~$x_i$).  $\epsilon$ either may be viewed at face
	value as a mathematical artifice or may be viewed as being
	equivalent to the multiple-scale assumption of variations occurring
	only on a large lateral length scale (large compared to the
	thickness of the fluid).  The two viewpoints provide exactly the
	same results.  In either case, let the lateral variations scale
	with the parameter $\epsilon$:
\begin{displaymath}
\D{x_i}{}
=\epsilon\D{x^\ast_i}{}\,,\quad
k_1=\epsilon k_1^{\ast}\,,\quad
k_2=\epsilon k_2^{\ast}\,,\quad
\kappa=\epsilon{\kappa}^{\ast}\,,   
\end{displaymath}
where ${\ast}$ denotes quantities which have been scaled
by~$\epsilon$.

	\item Second, the presumed small gravitational forcing is
	treated as a perturbing ``nonlinear'' effect by introducing the
	parameter~$\beta$ such that the gravity number $\gr =\beta^2$\,.

	\item Third, as introduced by Roberts~\cite{Roberts96b}, we also
	modify the tangential stress condition~(\ref{bcon4}) on free
	surface, using a parameter~$\gamma$: at $\gamma=0$ the lateral
	shear mode of slowest decay actually becomes a marginally stable
	mode; whereas at $\gamma=1$ the modification vanishes to restore
	the physical stress-free boundary condition~(\ref{bcon4}).  The
	modification to~(\ref{nbcon3}) is necessary to create the necessary
	three modes of the centre manifold model.  Subsequently, evaluating
	at $\gamma=1$ removes the modification to obtain a model for the
	physically correct dynamics.
\end{enumerate}
Now rewrite the governing fluid equations according to the above
tricks.  For convenience, we drop the ``$\ast$'' superscript on all
re-scaled variables hereafter.
Equations~(\ref{curequ1}--\ref{curequ2}) become
\begin{eqnarray}&&
\epsilon\D{x_1}{}(h_2 u_1)+\epsilon\D{x_2}{}(h_1 u_2)+
\D{y}{}(h_1h_2 v)=0\,,
\label{ncon}\\&&
\re\left\{\D{t}{{\bfa u}}
+{\bfa e}_1\left[\epsilon\frac{u_1}{h_1}\D{x_1}{u_1}+\epsilon\frac{u_2}{h_2}\D{
x_2}{u_1}+
v\D{
y}{u_1}+\epsilon\frac{u_2}{h_1h_2}\left(u_1\D{
x_2}{h_1}-u_2\D{x_1}{h_2}\right)
\right.\right.\nonumber\\&&\left.\left.\qquad{}
-\epsilon m_1k_1\frac{vu_1}{h_1}\right]
+{\bfa e}_2\left[\epsilon\frac{u_1}{h_1}\D{x_1}{u_2}+\epsilon\frac{u_2}{h_2}\D{
x_2}{u_2}+v\D{
y}{u_2}
\right.\right.\nonumber\\&&\left.\left.\qquad{}
+\epsilon\frac{u_1}{h_1h_2}\left(u_2\D{
x_1}{h_2}-u_1\D{x_2}{h_1}\right)-\epsilon 
m_2k_2\frac{vu_2}{h_2}\right]
\right.\nonumber\\&&\left.\quad{}
+{\bfa e}_3\left[\epsilon\frac{u_1}{h_1}\D{x_1}{v}
+\epsilon\frac{u_2}{h_2}\D{
x_2}{v}+   
v\D{y}{v}+\epsilon 
m_1k_1\frac{{u_1}^2}{h_1}+\epsilon 
m_2k_2\frac{{u_2}^2}{h_2}\right]\right\}
\nonumber\\&=&
-\epsilon\frac{{\bfa e}_1}{h_1}\D{x_1}{p}
-\epsilon\frac{{\bfa e}_2}{h_2}\D{x_2}{p}-{\bfa 
e}_3\D y{p} 
+\frac{1}{h_2}\left[\epsilon\D{
x_2}{\omega_3}-\D{y}{\omega_2}\right]{\bfa 
e}_1
\nonumber\\&&{}
+\frac{1}{h_1}\left[\D{
y}{\omega_1}-\epsilon\D{x_1}{\omega_3}\right]{\bfa e}_2+
\frac{\epsilon}{h_1h_2}\left[\D{
x_1}{\omega_2}-\D{x_2}{\omega_1}\right]{\bfa e}_3
+\beta^2\hat{\bfa g}\,,
\label{nnst}
\end{eqnarray}
where the scale factors are $h_i=m_i(1-\epsilon k_i y)$\,, and 
the components of the vorticity are
\begin{eqnarray*}
\omega_{i'}=\frac{(-1)^i}{h_i}\left[\epsilon\D{x_i}v
-\D{y}{(h_iu_i)}\right],\quad
\omega_3=\frac{\epsilon}{h_1h_2}\left[\D{x_1}{(h_2u_2)}
-\D{x_2}{(h_1u_1)}\right]. 
\end{eqnarray*}
The boundary conditions~(\ref{bcon1}--\ref{bcon4}) become
\begin{equation} 
{\bfa u}={\bfa 0}\quad\mbox{on $y=0$}\,, 
\label{nbcon1}
\end{equation}
\begin{equation}
\D{
t}{\eta}=v-\epsilon\frac{u_1}{\tilde{h}_1}\D{x_1}{\eta}
-\epsilon\frac{u_2}{\tilde{h}_2}\D{x_2}{\eta}\quad\mbox{on $y=\eta$}\,, 
\label{nbcon2}
\end{equation}
\begin{equation}
\tilde{\bfa t}_i\cdot\tilde{\bfa {\tau}}\cdot\tilde{\bfa n}= 
(1-\gamma)\frac{m_im_1m_2u_i}{\eta l_i l}
\quad\mbox{on $y=\eta$}\,, 
\label{nbcon3}
\end{equation}
\begin{equation}
\tilde{\bfa n}\cdot\tilde{\bfa {\tau}}\cdot\tilde{\bfa 
n}=\tilde p+\we\tilde{\kappa},
\label{nbcon4}
\end{equation}
where 
\begin{displaymath}
l_i=\sqrt{{\tilde{h}_i}^2+{\epsilon^2\eta_{x_i}^2}}\,,\quad
l=\sqrt{(\epsilon\tilde{h}_2\eta_{x_1})^2+(\epsilon\tilde{h}_1\eta_{x_2})^2
+(\tilde{h}_1\tilde{h}_2)^2}\,,
\end{displaymath}
and the unit tangent vectors~$\tilde{\bfa t}_i$ and unit normal vector 
$\tilde{\bfa n}$ are
\begin{displaymath}
\tilde{\bfa t}_i=(\tilde{h}_i{\bfa e}_i+\epsilon\eta_{x_i}{\bfa 
e}_3)/{l_i}\,,
\quad
\tilde{\bfa n}=(-\epsilon\tilde{h}_2\eta_{x_1}{\bfa 
e}_1-\epsilon\tilde{h}_1\eta_{x_2}{\bfa e}_2
+\tilde{h}_1\tilde{h}_2{\bfa e}_3)/l\,. 
\end{displaymath}
The asymptotic expressions for the deviatoric stress~$\tilde{\bfa \tau}$ 
on the free surface are
\begin{eqnarray}
\tilde{\tau}_{ii} & = &2\epsilon\left(\frac{1}{m_1}\D{x_i}{u_i}
+\frac{h_{i,i'}}{m_im_i'}u_i-k_iv\right)+\Ord{\epsilon^2}\,,\nonumber
\\
\tilde{\tau}_{12} & = &\epsilon\left(\frac{1}{m_2}\D{x_2}{u_1}
+\frac{1}{m_1}\D{x_1}{u_2}
-\frac{h_{1,2}}{m_1m_2}u_1
-\frac{h_{2,1}}{m_1m_2}u_2\right)
+\Ord{\epsilon^2}\,, 
\\
\tilde{\tau}_{i3} & = &\D{y}{u_i}+{\epsilon}
\left({k_i}u_i+\frac{1}{m_i}\D{
x_i}{v}\right)+\Ord{\epsilon^2}\,,\nonumber\\
\tilde{\tau}_{33} & = & 2\D{y}{v}\,,\nonumber 
\end{eqnarray}
and the mean curvature of the free surface~$\tilde{\kappa}$, expanded 
in powers of~$\epsilon$, is
\begin{displaymath}
\tilde{\kappa}=\epsilon\kappa+\epsilon^2[\nabla^2{\eta}-\kappa_2\eta]
+\Ord{\epsilon^3}\,, 
\end{displaymath}
where $\nabla^2{\eta}$ is the same as that in~(\ref{surfna}) and 
$\kappa_2=k_1^2+k_2^2$\,.  

The tangential stress boundary condition~(\ref{nbcon3}) has been 
modified by the introduction of the artificial parameter~$\gamma$.
The physically correct boundary condition is recovered when $\gamma=1$\,.
But when $\gamma=0$ the boundary condition~(\ref{nbcon3}) linearises 
to
\begin{displaymath}
    \D y{u_i}=\frac{u_i}{\eta}\,,\quad\mbox{on $y=\eta$\,,}
\end{displaymath}
which gives two neutral horizontal shear modes, $u_i\propto y$\,.
The above equations have generalised the physical equations by 
introducing the extra parameters~$\epsilon$, $\gamma$ and~$\beta$.
Then by adjoining the trivial equations
\begin{equation}
\D{t}{\epsilon}=0\,, 
\quad \D{t}{\gamma}=0 
\quad\mbox{and}\quad \D{t}{\beta}=0\,,
\end{equation}
we obtain a new dynamical system in the variables~${\bfa u}$, $\eta$,
$p$, $\epsilon$, $\gamma$ and~$\beta$.  The original system will be
recovered by setting~$\epsilon=1$, $\gamma=1$ and~$\beta=\sqrt{\gr
}$\,.  However, the two systems are quite different from the view of
centre manifold theory.  The theory now treats all terms that are
multiplied by the three introduced parameters as nonlinear perturbing
effects on the system.  So the dynamics we describe will be suitable
only when there are slow lateral variations in~$x_i$ of the curvatures
of the substrate, of~${\bfa u}$, $p$ and~$\eta$, small $\epsilon$, and
a relatively weak gravitational forcing on the system, small~$\gr$.  In
\S\ref{ldm3d} we argue that evaluating at $\gamma=1$ is sound, and
towards the end of \S\ref{ldm3d} we give evidence that lateral
variations are slow enough if their logarithmic derivative is less
than~$1.9/\eta$\,.  We now proceed to use centre manifold techniques to
develop a model of the dynamics.
 
A linear picture of the dynamics is fundamental to the application of
centre manifold techniques to derive a low-dimensional model.  The
linear part of system~(\ref{ncon}--\ref{nnst}), that is omitting all
terms multiplied by a small parameter $\epsilon$, $\gamma$~or~$\beta$,
is
\begin{eqnarray}
\D{y}{v}&=&0\,,
\label{lin1}
\\
\re\D{t}{{\bfa u}}+{\bfa e}_3 
\D{y}{p}
-\DD{y}{{\bfa u}}&=&0\,,
\label{lin2}
\end{eqnarray}
with the boundary conditions~(\ref{nbcon1}--\ref{nbcon4}) linearised 
to
\begin{eqnarray}
{\bfa u}&=&{\bfa 0}\quad\mbox{on $y=0$}\,,\nonumber \\
\D{t}{\eta}-v&=&0\quad\mbox{on 
$y=\eta$}\,,\nonumber \\ 
\D{y}{u_i}-\frac{u_i}{\eta}&=&0\quad\mbox{on 
$y=\eta$}\,,
\label{linbcon}
\\
2\D{y}{v}-p&=&0\quad\mbox{on 
$y=\eta$}\,.\nonumber 
\end{eqnarray}
Note that there are no curvature nor lateral variations in the above 
linear equations, such variations are not included in the leading 
order approximation to the physical system.
The linear dynamical system has three types of solutions: first, a 
motionless film of constant thickness~$u=v=p=0$\,, 
$\eta=\mbox{constant}$; second, the family of lateral shear modes 
$u_i\propto \sin(\chi y/\eta)\exp(\lambda t)$ where
\begin{equation}
\lambda=-\frac{\chi^2}{\re\eta^2}\,,\quad\mbox{such that}\quad 
\chi=\tan\chi\,; 
\label{linso}
\end{equation}
and third, the trivial~$\epsilon$, $\gamma$ and~$\beta$ being 
independently constant.
Thus, the six modes corresponding to zero eigenvalues, the so-called 
critical modes, are the four modes with~$\eta$, $\epsilon$, $\gamma$ 
and~$\beta$ arbitrarily constant, and the two modes with $u_i \propto 
y$ (obtained in the limit $\chi\to0$).
All other modes correspond to negative eigenvalues given 
by~(\ref{linso})---they are damped by viscosity.
Consequently the centre manifold model which we create has six modes: 
three corresponding to critical physical modes; and three 
corresponding to trivial parameter modes.

Centre manifold techniques are justifiably applied to infinite dimensional
dynamical systems whose linearisation has only non-positive eigenvalues
and such that the nonlinear perturbation terms in the system are smooth
and bounded, see Gallay~\cite{Gallay93}.  The perturbation terms in
system~(\ref{ncon}--\ref{nnst}), involving spatial derivatives, are
strictly speaking unbounded so the rigorous theory does not apply.
Nonetheless, by restricting attention to slowly varying solutions only,
the derivatives remain bounded and the resulting model is expected to
be accurate~\cite{Roberts88a}.  With this proviso, a low-dimensional
model of the system may be derived using centre manifold theory.

Denote the physical fields by ${\bfa v}(t)= (\eta, u_1, u_2, v,p)$\,.
Centre manifold theory guarantees that there exist functions {\bfa V} 
and {\bfa G} of the critical modes where the critical modes evolve in 
time, that is
\begin{equation}
{\bfa v}(t)={\bfa V}(\eta,\bar{u}_1,\bar{u}_2),\quad\mbox{such 
that}\quad \D{t}{}\left[\begin{array}{c} \eta \\ 
\bar{u}_1 \\ \bar{u}_2\end{array} \right]
={\bfa G}(\eta,\bar{u}_1,\bar{u}_2),
\label{trans}
\end{equation}
where there is implicit dependence upon the parameters 
$(\epsilon,\gamma,\beta)$\,, which are treated as small constants, and 
where~$\bar{u}_i$ are depth-averaged lateral velocities defined 
precisely as
\begin{equation}
    \bu _i=\frac{1}{\eta}\int_0^\eta u_i(1-k_{i'}y)\,dy\,.
    \label{eq:baru}
\end{equation}  
This definition ensures that the fluid flux over a point on the
substrate is simply~$\eta\bbu$\,.  We proceed to find functions~${\bfa
V}$ and~${\bfa G}$ such that ${\bfa v}(t)$ as described
by~(\ref{trans}) are actual solutions of the physical
equations~(\ref{ncon}--\ref{nnst}) satisfying boundary
conditions~(\ref{nbcon1}--\ref{nbcon4}).  We calculate~${\bfa V}$
and~${\bfa G}$ by an iteration using computer
algebra~\cite{Roberts96a}, see Appendix~\ref{3dp}.  In outline, suppose
that an approximation~$\tilde{\bfa V}$ and~$\tilde{\bfa G}$ has been
found, and let~${\bfa V'}$ and~${\bfa G'}$ denote corrections we seek
to improve~$\tilde{\bfa V}$ and~$\tilde{\bfa G}$\,.  Substituting
\begin{displaymath}
{\bfa v}=\tilde{\bfa V}+{\bfa V'},\quad\D{
t}{}\left[\begin{array}{c} \eta \\ \bar{u}_1 \\ \bar{u}_2\end{array} 
\right] =\tilde{\bfa G}+{\bfa G'} 
\end{displaymath}
into~(\ref{ncon}--\ref{nnst}) and its boundary conditions then 
rearranging, dropping products of corrections, and using the linear 
approximation wherever factors multiply corrections 
(see~\cite{Roberts96a} for more details), we obtain a system of linear 
equations for the corrections.
The resulting system of equations is in the homological form
\begin{equation}
{\cal L}{\bfa V'}+A{\bfa G'}=\tilde{\bfa R}\,,
\label{eq:homo}
\end{equation}
where~${\cal L}$ is the linear operator on the left hand side of
system~(\ref{lin1}--\ref{linbcon}), $A$ is a matrix, and~$\tilde{\bfa
R}$ is the residual of the governing equations (\ref{ncon}--\ref{nnst})
and boundary conditions~(\ref{nbcon1}--\ref{nbcon4}) using the reigning
approximations~$\tilde{\bfa V}$ and~$\tilde{\bfa G}$\,.  The procedure
for solving the homological equation~(\ref{eq:homo}) is as follows:
first, choose~${\bfa G'}$ such that $\tilde{\bfa R}-A{\bfa G'}$ is in
the range of~${\cal L}$; second, solve ${\cal L}{\bfa
V'}=\mbox{r.h.s.}$ making the solution satisfy the boundary
conditions~(\ref{nbcon1}--\ref{nbcon4}) and the
definitions~(\ref{eq:baru}).  Then regard $\tilde{\bfa V}+{\bfa V'}$
and $\tilde{\bfa G}+{\bfa G'}$ as the new approximation~$\tilde{\bfa
V}$ and~$\tilde{\bfa G}$ respectively.  Repeat the iteration until
satisfied with the approximation.  The ultimate purpose is to make the
residual~$\tilde{\bfa R}$ become zero to the required order of error,
then the Approximation Theorem~\cite{Carr81} in centre manifold theory
assures us that the low-dimensional model has the same order of error.
The computer algebra program listed in Appendix~\ref{3dp} performs the
computations.

\section{The high order model of film flow}
\label{ldm3d}
The computer algebra program listed in Appendix~\ref{3dp} gives the
physical fields of slowly varyingthin film fluid flow, and also
describes the evolution thereon as a set of coupled partial
differential equations for the evolution of the film thickness~$\eta$
and the averaged lateral velocities~$\bbu$.

Computing to low order in the small parameters, it gives the following 
fields in terms of the parameters and a scaled normal 
coordinate~$Y=y/\eta$:
\begin{eqnarray}
p&=&-\epsilon\we\kappa-\epsilon^2\we{\nabla}^2\eta
   +\epsilon{\eta}^{-1}\nabla\eta\cdot\bbu(Y-1)(2+\gamma/2)
   -\epsilon^2\we\eta\kappa_2
\nonumber\\
 & &{}+\epsilon\nabla\cdot\bbu\left(\gamma (Y-3)/2
-2(Y+1)\right)+\gr\eta g_n(Y-1)
\nonumber\\
& &{}+\Ord{\epsilon^3+\bar{u}^3+\beta^3,\gamma^2}\,;
\label{pex}\\
{\bfa u}&=&\bbu\left(2Y-\gamma (Y^3-Y/2)\right)
         +\eta^2\left(\epsilon^2\we\nabla\kappa+\gr{\bfa g}_s\right)
	 \left(-\frac{3}{80}\gamma Y^5
         \right.
\nonumber\\
        & &\left.\quad{} +\frac{23}{240}\gamma Y^3
             -\frac{17}{480}\gamma Y+\frac{1}{4}Y^3
          -\frac{1}{2}Y^2+\frac{5}{24}Y\right)
\nonumber\\
        & &{}+\epsilon\eta\kappa\bbu\left(\frac{3}{20}\gamma 
        Y^5-\frac{1}{2}\gamma Y^4
             -\frac{23}{60}\gamma Y^3
          +\frac{1}{4}\gamma Y^2+\frac{13}{120}\gamma Y\right.
\nonumber\\
        & &\left.\quad{}-\frac{1}{2}Y^3+Y^2
             +\frac{11}{12}Y\right)
          +\epsilon\eta{\bfa K}\cdot\bbu\left(\frac{3}{10}\gamma 
          Y^5+\frac{17}{60}\gamma Y^3
\right.
\nonumber\\
       & &\left.\quad{}-\frac{19}{60}\gamma Y-Y^3
             -\frac{5}{6}Y\right)
           +\Ord{\epsilon^3+\bar{u}^3+\beta^3,\gamma^2}\,; 
\label{uex}\\
v&=&\epsilon\nabla\eta\cdot\bbu\left(\gamma (-3Y^4+Y^2)/4
   +Y^2\right)
\nonumber\\
 & &{}+\epsilon\eta\nabla\cdot\bbu\left(\gamma (Y^4-Y^2)/4
   -Y^2\right)
\nonumber\\
 & &{}+\Ord{\epsilon^3+\bar{u}^3+\beta^3,\gamma^2}\,; 
\label{vex}
\end{eqnarray}
where $g_n$ is the component of gravity in the direction normal to the
substrate.  The error terms in these expressions involve
$\bar{u}=\|\bar{\bfa u}\|$\,, and then
$\Ord{\epsilon^p+\bar{u}^q+\beta^m,\gamma^n}$ is used to denote
terms~$s$ for which either $s/(\epsilon^p+\bar{u}^q+\beta^m)$ is
bounded as $(\epsilon,\bar{u},\beta)\rightarrow\bfa0$\,, or
$s/\gamma^n$ is bounded as $\gamma\rightarrow0$\,.  The corresponding
evolution to this order of accuracy is
\begin{eqnarray}
\D{t}{\eta}&=&-\epsilon{\nabla}\cdot(\eta\bbu) 
+\Ord{\epsilon^3+\bar{u}^3+\beta^3,\gamma^2},
\label{firstre}
\\
\nonumber
\\
\re\D{t}{\bbu}&=& 
         (\epsilon^2\we{\nabla}\kappa+\gr{\bfa g}_s)
	 \left(\frac{3}{4}+\frac{1}{10}\gamma\right) 
         -\epsilon\eta^{-1}{\bfa K}\cdot\bar{\bfa 
u}\left(3-\frac{6}{5}\gamma\right)
\nonumber\\
       & &{}-\epsilon{\eta}^{-1}\kappa\bar{\bfa 
u}\left(\frac{3}{2}-\frac{3}{5}\gamma\right)
         -3\eta^{-2}\bbu\gamma
    \nonumber\\
& &{}+\Ord{\epsilon^3+\bar{u}^3+\beta^3,\gamma^2}\,.   
\label{sample}
\end{eqnarray}
Throughout~${\bfa K}$ is the curvature tensor which in the special 
coordinate system chosen to fit the substrate takes the diagonal form
\begin{equation}
{\bfa K}=\left(\begin{array}{cc}
              k_1 & 0\\
               0  & k_2
                 \end{array}\right)\,. 
\label{curten}
\end{equation}
Although derived in the special coordinate system, the above and later 
results are all written in a coordinate free form.
The differential operators that appear are those of the substrate.
In the special orthogonal coordinate system they involve the substrate 
scale factors~$m_i$ as in the Laplacian~(\ref{surfna}).

To recover the original model, we need to set~$\gamma=1$ so
that~(\ref{nbcon3}) reverts to the physically correct boundary
condition.  In the above asymptotic expansions every coefficient is a
series in~$\gamma$, and the ratios of the coefficients of
$\gamma^{n-1}$ to~$\gamma^n$ in all such series appear to be greater
than about~$1.5$ for~$n>2$ from further calculation.  That is, the
radii of convergence of the various series' in~$\gamma$ are greater
than about~$1.5$\,.  Table~\ref{coef} shows the coefficients of the
$\gamma$~series of some terms in a higher order version of the
low-dimensional model~(\ref{sample}).  Evidently the convergence of at
least these series' is very good---we may expect five decimal place
accuracy from the shown terms and similar for the other coefficients.
A similar argument on the convergence of such series is reported
in~\cite{Roberts96b,Roberts94c}.
\begin{table}
\begin{center}
\begin{tabular}{|c|rrr|}\hline
 & $\bbu/\eta^2$ & $\bfa g_s$ & $\bbu\kappa/\eta$ \\
\hline
$1$       &$0       $&$+0.75000$&$-1.50000$\\
$\gamma$  &$-3.00000$&$+0.10000$&$+0.60000$\\
$\gamma^2$&$+0.60000$&$-0.03286$&$-0.10286$\\
$\gamma^3$&$-0.06857$&$+0.00571$&$0       $\\
$\gamma^4$&$   0    $&$-0.00032$&$+0.00321$\\
$\gamma^5$&$+0.00128$&$-0.00009$&$-0.00024$\\
$\gamma^6$&$-0.00008$&$+0.00002$&$-0.00014$\\
$\gamma^7$&$-0.00004$&$+0.00000$&$+0.00003$\\
\hline 
\end{tabular}
\caption{some higher order terms in the series expansions in~$\gamma$ 
of selected coefficients in the low-dimensional model~(\ref{ldm3d2}) 
showing that these expansions are effectively summed at~$\gamma=1$\,.}
\label{coef}
\end{center}
\end{table}
Hence we justifiably substitute~$\gamma=1$ into the series of every
coefficient to obtain the physical model.  Here we generally calculate
every coefficient in the evolution from the terms in the series up to
and including those of order~$\gamma^7$.  We also now set~$\epsilon=1$
to recover the unscaled model of the original dynamics.  With higher
order corrections in~$\gamma$ the low-dimensional model~(\ref{sample})
then becomes the model~(\ref{my2}) discussed briefly in the
Introduction.

\begin{figure}[tbp]
    \centering
    \includegraphics{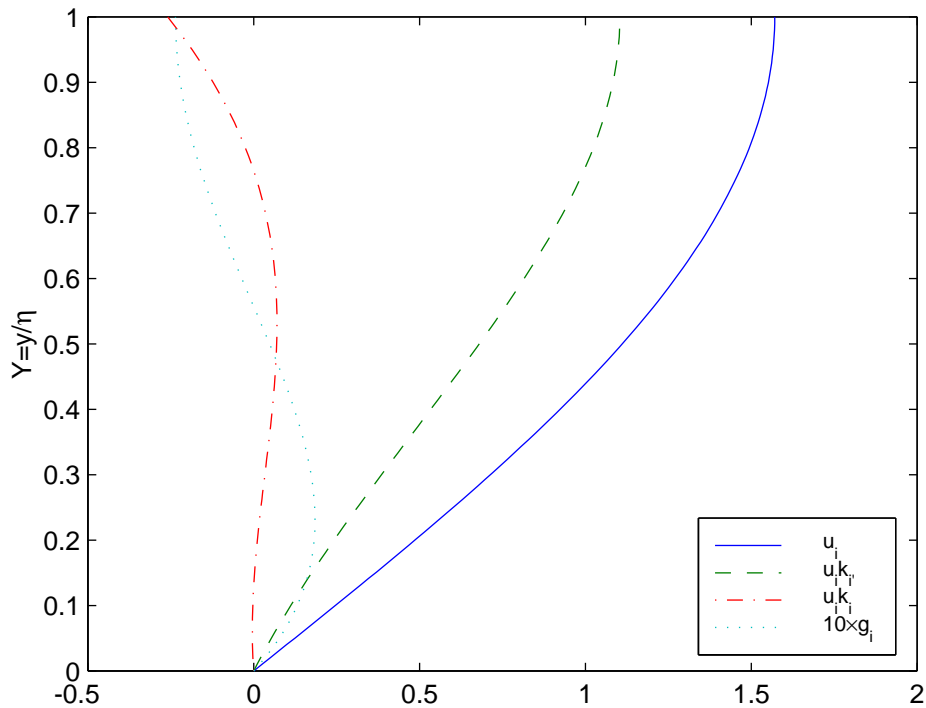}
    \caption{the normal structure of the lateral velocity 
    field~(\ref{uex}): solid, component $\propto\bbu$; dashed, 
    component $\propto k_{i'}u_i$; dot-dashed, component $\propto 
    k_iu_i$; and dotted, $10\times$ component $\propto 
    \we\nabla\kappa+\gr \bfa g_s$\,.}
    \label{fig:uprof}
\end{figure}

The low order expressions~(\ref{pex}--\ref{vex}), when
$\gamma=\epsilon=1$\,, give approximations to the physical state of the
fluid flow corresponding to a given~$\eta$, $\bu_1$ and~$\bu_2$\,.
Higher order expressions for the normal structure of the lateral
velocity~$\bfa u$ are plotted in Figure~\ref{fig:uprof}.  The solid
curve shows the fundamental structure of the lateral velocity in the
normal direction; qualitatively it is dominantly parabolic, but it
differs in detail.  It is indistinguishable from the trigonometric
$\frac{\pi}{2}\sin(\pi Y/2)$ expected from the correct linear
problem~\cite{Roberts94c}, and thus is slightly faster at the free
surface than the parabolic profile with the same flux.  The dashed
curve shows that in order to maintain the flux~$\bbu$ the flow~$u_i(Y)$
along a trough, $k_{i'}>0$\,, has to be proportionally faster.  The
dot-dashed curve shows that flow curving upwards, around an internal
corner, is slower at the free surface and conversely faster for flow
around an external corner; part of this effect could be attributed to
solid body rotation being the dissipation free mode for turning a
corner.  Lastly, the dotted curve, exaggerated by a factor of ten,
shows the very small adjustment made to the profile when the flow is
driven by gravity or lateral pressure gradients---observe the velocity
at the free-surface will decrease slightly so that when lateral forces
exactly balance the drag on the substrate the profile will be then the
familiar parabolic Pouiseille flow.  These show how just some of the
physical processes affect the details of the physical fields and thus
indirectly influence the evolution.

The shear stress on the substrate is of interest:
\begin{eqnarray}
    \bfa\tau_y &=& 2.467\frac{\bbu}{\eta}
    +0.1775\,\eta\left(\we\nabla\kappa +\gr{\bfa g}_s\right)
    +(\kappa\bfa I-3.609\,\bfa K)\cdot\bbu
    \nonumber\\&&{}
    +\Ord{\partial_x^3+\bar{u}^3+\gr^{3/2}}\,.
    \label{eq:subshear}
\end{eqnarray}
The first term is just the viscous drag on a flat substrate.
The next is the enhanced stress transmitted to the substrate when the 
fluid is driven by a body force or pressure gradients, equivalently.
The third and last term accounts for the effects of curvature on the 
velocity field affecting the velocity profile near the bed.

\paragraph{Higher order model.} With computer algebra we readily
compute a more accurate, more comprehensive higher order model.
Atherton \& Homsy~\cite{Atherton73} and Lange~\cite{Lange99} have
similarly considered high order models of thin film flows obtained via
computer algebra but only in the lubrication approximation.  Computing
to the next order in spatial gradients~$\epsilon$, velocity
field~$\bu$, and gravitational forcing~$\beta$, the model is written as
follows:\footnote{Some of the constants that appear here are
tentatively identified: $1.0779= (\pi^2+16)/24$\,,
$0.4891=(\pi^2-4)/12$\,, $2.4099=\pi^2/7+1$ and perhaps
$4.0930=(8\pi^2+7)/21$\,.}
\begin{eqnarray}
\D{t}{\zeta}&=&-{\nabla}\cdot(\eta\bbu)\,,
\label{ldm3d1}
\\
\re\D{t}{\bbu}
         &=&
-\left[\frac{\pi^2}{4}\frac{\bbu}{\eta^2}
+(2{\bfa K}+\kappa{\bfa I})\cdot\frac{\bbu}{\eta}
\right.\nonumber\\
(\mbox{\sf drag})
&&\left.\vphantom{\frac{\pi^2}{4}}\quad{}
+(3.2974\,{\bfa K}\cdot{\bfa K}
-1.1080\,\kappa{\bfa K} 
+0.6487\,\kappa_2{\bfa I})\cdot\bbu\right] 
\nonumber\\ 
(\mbox{\sf tension})    
&&{}
+\we\left[\frac{\pi^2}{12}\nabla(\kappa+\eta\kappa_2+\nabla^2\eta) 
+1.0779\,\eta{\bfa K}\cdot\nabla{\kappa} 
-0.4891\,\eta\kappa\nabla\kappa\right]
\nonumber\\
(\mbox{\sf gravity})
&&{}
+\gr \left[\frac{\pi^2}{12}({\bfa g}_s+g_n\nabla\eta)
+0.2554\,\eta{\bfa K}\cdot{\bfa g}_s
-0.4891\,\eta{\kappa}{\bfa g}_s) \right] 
\nonumber\\
(\mbox{\sf advect})
&&{}
-\re\left[
1.3464\,\bbu\cdot\nabla\bbu
+\left(0.1483\,\bbu\cdot{\nabla\eta}/\eta
+0.1577\,\nabla\cdot\bbu \right)\bbu\right]
\nonumber\\ 
(\mbox{\sf viscous})
&&{}
+\frac{4.0930}{{\eta}^{0.8348}}\nabla\left[\eta^{0.4886}\nabla\cdot
       \left({\eta}^{0.3461}{\bbu}\right)\right] 
\nonumber\\&&\quad{}
-\frac{1}{\eta^{0.4377}}\nabla\times\left[
 \frac{1}{\eta^{1.0623}}\nabla\times\left(
 \eta^{3/2}{\bbu} \right)\right]
\nonumber\\ 
&&{}
+0.9377\frac{1}{\eta}\nabla\eta\times(\nabla\times{\bbu})
-2.4099\frac{\bbu}{\eta^{0.8299}}\nabla^2\left(\eta^{0.8299}\right)
\nonumber\\&&{}
+\Ord{{\nabla}^4+\bar{u}^4+\gr ^2}\,,
\label{ldm3d2}
\end{eqnarray}
where the differential operators are those of the substrate coordinate 
system, noting in particular that~\cite[p599]{Batchelor79}
\begin{eqnarray*}
    \nabla\times\bbu & = & \bfa e_3\frac{1}{m_1m_2}\left[ 
    \D{x_1}{(m_2\bu _2)}-\D{x_2}{(m_1\bu _1)} \right]\,,  \\
    \nabla\times(\bfa e_3\omega) & = & \bfa 
    e_1\frac{1}{m_2}\D{x_2}\omega-\bfa 
    e_2\frac{1}{m_1}\D{x_1}\omega\,,  \\
    \bbu\cdot\nabla\bbu & = & \bfa e_1\left[ 
    \bbu\cdot\nabla \bu _1 +\frac{\bu _2}{m_1m_2}\left( 
    \bu _1\D{x_2}{m_1}-\bu _2\D{x_1}{m_2} \right) \right] 
    \\&&{}
    +\bfa e_2\left[ \bbu\cdot\nabla \bu _2 +\frac{\bu _1}{m_1m_2}\left( 
    \bu _2\D{x_1}{m_2}-\bu _1\D{x_2}{m_1} \right) \right]\,.
\end{eqnarray*}
Observe that fluid is conserved by~(\ref{ldm3d1}).  In the above
model~(\ref{ldm3d2}) for the average velocity field we identify the
apparent physical source of the terms in the various lines by the
cryptic words to the left of the lines.  Generally the viscous drag on
the bed, surface tension forces and gravitational forcing show some
subtle effects of the curvature of the substrate.  In faster flows of
higher Reynolds number, the most usually modelled part of the advection
terms, the self-advection term~$\bar{\bfa u}\cdot\nabla\bbu$\,, has the
definite coefficient~$1.3464$\,.  But note that some of the
self-advection is also encompassed within the~$(\nabla\cdot\bbu)\bbu$
term.  This modelling settles, see for
example~\cite[Eqn.(19)]{Prokopiou91b} or~\cite[Eqn.(23)]{Prokopiou91},
the correct theoretical value for this and other coefficients.  The
lateral damping via viscosity seems most natural to express in a mixed
form involving both the general grad-div operator and the curl-curl
operator (recall the vector identity $\nabla^2\bfa
u=\nabla(\nabla\cdot\bfa u)-\nabla\times(\nabla\times\bfa u)$).  The
involvement of fractional powers of the film thickness within the scope
of these operators is a convenient way to reduce the number of terms
within the equation; as yet we have not discerned any interesting
physical significance to this arrangement.  You may truncate the above
model in a variety of consistent ways depending upon the parameter
regimes of the application you are considering.

\paragraph{Lubrication models} such as~(\ref{roy}) may be derived
from~(\ref{ldm3d1}--\ref{ldm3d2}).  Obtain simple low order accurate
models simply by balancing the drag terms,
dominantly~$(\pi^2/4)\bbu/\eta^2$\,, against the driving forces
expressed by the surface tension and gravity terms.  This then
expresses the average velocity field~$\bbu$ as a function of the film
thickness~$\eta$.  This is substituted into the conservation of fluid
equation~(\ref{ldm3d1}) to form a lubrication model.  Higher order,
more sophisticated models are formed by then taking into account the
consequent time dependence of the weaker previously neglected terms
in~(\ref{ldm3d2}).  Any resultant lubrication model is correct because
rational mathematical modelling is transitive: a coarser model of a
model of some dynamics is the same as the coarser model derived
directly.
 

\paragraph{Slower flow} occurs when in a specific application the
velocity field is predominantly driven by surface tension acting
because of curvature gradients, whence $\bu=\Ord{\nabla\kappa}$\,.  The
lateral velocities are significantly damped by viscous drag on the
substrate.  In this case truncate~(\ref{ldm3d2}) to
\begin{eqnarray}
\re\D{t}{\bbu}
         &=&
-\left[\frac{\pi^2}{4}\frac{\bbu}{\eta^2}
+(2{\bfa K}+\kappa{\bfa I})\cdot\frac{\bbu}{\eta}
\right] 
\nonumber\\ 
&&{}
+\we\left[\frac{\pi^2}{12}\nabla\tilde\kappa 
+1.0779\,\eta{\bfa K}\cdot\nabla{\kappa} 
-0.4891\,\eta\kappa\nabla\kappa \right]
\nonumber\\
&&{}
+\gr \left[\frac{\pi^2}{12}({\bfa g}_s+g_n\nabla\eta)
+0.2554\,\eta{\bfa K}\cdot{\bfa g}_s
-0.4891\,\eta{\kappa}{\bfa g}_s) \right] 
\nonumber\\&&{}
+\Ord{{\nabla}^4+\bar{u}^2+\gr ^2}\,,
\label{eq:utrunc}
\end{eqnarray}
That is, you may adjust the dynamical model~(\ref{ldm3d2}) to suit a
particular application by choosing an appropriate consistent
truncation.

\paragraph{Order one gradients are encompassed} by the
model~(\ref{ldm3d2}).  `Long wave' models such as~(\ref{ldm3d2})
and~(\ref{eq:utrunc}) are based upon the assumption that the lateral
spatial gradients are small.  We here quantify what a `small gradient'
means in this context following similar arguments for the Taylor model
of shear dispersion~\cite{Mercer90, Watt94b}.  We modify a simpler
version of the computer algebra derivation of Appendix~\ref{3dp} to
find the centre manifold model of the \emph{linear} dynamics about a
stationary constant thickness fluid with surface tension but no
gravity: $\eta=\eta_0+\alpha \eta'+\Ord{\alpha^2}$ and $\bfa u=\alpha
{\bfa u}'+\Ord{\alpha^2}$ where\footnote{The coefficients in the linear
model~(\ref{eq:holin}) come from evaluating at $\gamma=1$ an expansion
with errors~$\Ord{\gamma^7}$\,.  The coefficients used here should be
accurate as discussed earlier and demonstrated in Table~\ref{coef}.}
\begin{eqnarray}
    \D t{\eta'}&=&-\eta_0\divv{\bbu}'+\Ord{\alpha}\,,\\
    \D t{{\bbu}'}&=&
\frac1{\eta_0^2}\left[
-2.47
+4.09\,\eta_0^2\nabla^2
+.734\,\eta_0^4\nabla^4
+.0611\,\eta_0^6\nabla^6
+.0223\,\eta_0^8\nabla^8
\right]{\bbu}'
\nonumber\\&&{}
+\frac\we{\eta_0^3}\left[
.822\,\eta_0^3\nabla^3
+.116\,\eta_0^5\nabla^5
+.00168\,\eta_0^7\nabla^7
+.00298\,\eta_0^9\nabla^9
\right]\eta'
\nonumber\\&&{}
+\Ord{\alpha,\nabla^{10}}\,.
\label{eq:holin}
\end{eqnarray}
\begin{figure}[tbp]
    \centering
    \includegraphics{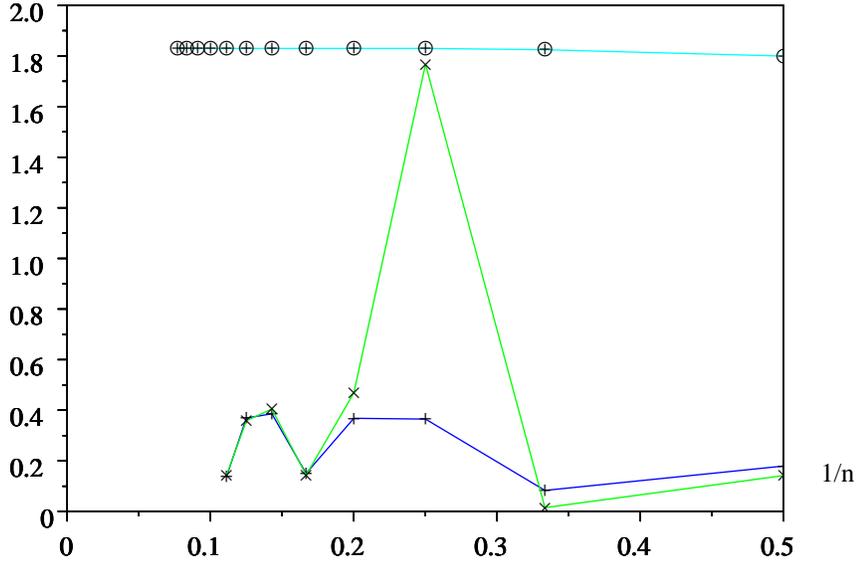}
  \caption{Domb-Sykes plot of the formal expansions~(\ref{eq:holin})
  extended to $\Ord{\nabla^{20}}$\,, $\times$~and~$+$, showing that the
  radius of convergence, $1/n\to 0$\,, may be roughly in the
  range~$1/0.15$ to~$1/0.35$\,.  The $\oplus$ Domb-Sykes plot is for
  the analogous lubrication model~(\ref{eq:holub}) showing its radius
  of convergence is~$1/1.83$\,.  }
    \label{fig:holin}
\end{figure}%
Evidently the linear dynamics are a formal expansion in
$\eta_0^2\nabla^2$\,.  This expansion will converge when the lateral
gradients are not too steep.  Suppose locally a solution has spatial
structure approximated by an exponential variation, say $\eta'\propto
e^{\nu x}$\,, then the above expansions become power series expansion
in~$\eta_0^2\nu^2$\,.  The Domb-Sykes plot~\cite{Mercer90} of the ratio
of successive coefficients in Figure~\ref{fig:holin} suggests that the
power series converges for $\eta_0^2\nu^2$ less than something roughly
in the range~$1/0.15$ to~$1/0.35$\,.  The constant sign of the
coefficients in~(\ref{eq:holin}) indicates the convergence limiting
singularity occurs for real steep gradients.  But the strong
period~3 oscillations in the Domb-Sykes plot of the ratio indicates a
complex conjugate pair of singularities occur at an angle of $\pm\pi/3$
to the real axis at nearly the same `distance'.  The generalisation of
the Domb-Sykes plot to cater for multiple comparable limiting
singularities~\cite{Watt94b} indicates that the three singularities are
at a distance about~$1/0.28$\,; that is, for any quantity~$w$, the
magnitude of the logarithmic derivative of the lateral structure
\begin{equation}
    |\nabla(\log w)|
    =\left|\frac{\nabla w}{w}\right|
    <\frac{1.9}{\eta}\,,
    \label{eq:logder}
\end{equation}
for the model to converge.  For example, apply this limit to the
surface thickness, $w=\eta$, to deduce that the steepness of the fluid
variations $|\nabla\eta|<1.9$\,, and that accurate approximation is
achieved for steepnesses significantly less than this rough limit.
Hence, steepnesses up to about one should be reasonably well
represented by those low order terms appearing in the
model~(\ref{ldm3d2}).

For interest, we also investigated the analogous but poorer spatial
resolution of the lubrication model~(\ref{roy}) of thin film
flow~\cite[e.g.]{Roy96}.  The analogous high order but linear model is
\begin{eqnarray}
    \D t{\eta'}&\!\!=\!\!&\we\left[
-.333\,\eta_0^3\nabla^4
-.6\,\eta_0^5\nabla^6
-1.09\,\eta_0^7\nabla^8
-2.00\,\eta_0^9\nabla^{10}
-3.67\,\eta_0^{11}\nabla^{12}
    \right]\eta' 
    \nonumber\\&&{}
    +\Ord{\alpha,\nabla^{14}}\,.
    \label{eq:holub}
\end{eqnarray}
Continuing this expansion to errors~$\Ord{\nabla^{30}}$ see in its
Domb-Sykes plot in Figure~\ref{fig:holin} that this power series
converges only for much less rapid variations than the
model~(\ref{ldm3d2}).  For example, the fluid thickness steepness
$|\nabla\eta|<0.74$\,, and so should be less than about a third, say,
in order for the usual first term in the lubrication model
$\eta_t=\third\we\divv(\eta^3\nabla^3\eta)$ to be reasonable.  Thus
expect the model~(\ref{ldm3d2}) developed here to resolve spatial
structure roughly three times as fine as a lubrication model.

\section{The model on various specific substrates}
\label{Sspec}

The model~(\ref{ldm3d1}--\ref{ldm3d2}) is quite sophisticated.  It is
not obvious how it will appear in any particular geometry.  Thus in
this section we record the model for four common substrate shapes:
flat, cylindrical, channel and spherical.  The models are given in
terms of elementary derivatives rather than vector operators for easier
use in specific problems.

\subsection{Flow on a flat substrate resolves a radial hydraulic jump}
\label{Sflat}

The simplest example is the flow on a flat substrate.  We: show 1D wave
transitions; simulate Faraday waves; explore divergence and vorticity
in the linearised dynamics; and lastly demonstrate that modelling the
inertia enables us to resolve hydraulic jumps in a radial flow.

On a flat substrate use a Cartesian coordinate system~$(x,y)$ and let
the mean lateral velocity~$\bbu$ have components~$\bu $ and~$\bv $
respectively (note that in this subsection $y$~is a tangential
coordinate, not a normal coordinate).  The substrate has scale
factors~$m_1=m_2=1$\,, and curvatures $k_1=k_2=0$\,.  The
model~(\ref{ldm3d1}--\ref{ldm3d2}) becomes, where~$g_n$ is the
direction cosine of gravity normal to the substrate into the fluid
layer and where subscripts on~$\eta$ denote partial derivatives,
\begin{eqnarray}
    \D t\eta & \approx & -\D x{(\eta\bu )}-\D y{(\eta 
    \bv )}\,,
    \label{eq:flath}  \\
    \re\D t{\bu } & \approx & 
    -\frac{\pi^2}{4}\frac{\bu }{\eta^2} 
    +\frac{\pi^2}{12}\left[ \gr (g_x +g_n\eta_x) 
    +\we\left(\eta_{xxx}+\eta_{xyy}\right) \right]
    \nonumber\\&&{}
    -\re\left[ 1.5041\,\bu \D x{\bu } 
    +1.3464\,{\bv }\D y{\bu } +0.1577\,{\bu }\D y{\bv } 
    \right.\nonumber\\&&\left.\quad{}
    +0.1483\frac{\bu }{\eta}\left(\bu {\eta}_x+{\bv }{\eta}_y\right)
    \right]
    \nonumber\\&&{}
    +\left[ 4.0930\DD x{\bu } +\DD y{\bu } 
    +3.0930\frac{\partial^2\bv }{\partial x\partial y}
    \right.\nonumber\\&&\left.\quad{}
    +4.8333\frac{\eta_x}{\eta}\D x{\bu } 
    +\frac{\eta_ y}{\eta}\D y{\bu } 
    +1.9167\frac{\eta_x}{\eta}\D y{\bv } 
    +1.9167\frac{h_ y}{\eta}\D x{\bv } 
    \right.\nonumber\\&&\left.\quad{}
    +\left( 
    -0.5033\frac{\eta_ y^2}{\eta^2} 
    -\frac{\eta_{y y}}{2\eta} 
    +0.1061\frac{\eta_x^2}{\eta^2} -0.5834\frac{\eta_{xx}}{\eta} 
    \right)\bu  
    \right.\nonumber\\&&\left.\quad{}
    +\left( 0.6094\frac{\eta_ y\eta_x}{\eta^2} 
    -0.0833\frac{\eta_{x y}}{\eta} \right)\bv  \right]
    \,,
    \label{eq:flatu}  \\
    \re\D t{\bv } & \approx & 
    -\frac{\pi^2}{4}\frac{\bv }{\eta^2} 
    +\frac{\pi^2}{12}\left[ \gr (g_ y +g_n\eta_ y) 
    +\we\left(\eta_{xxy}+\eta_{yyy}\right) \right]
    \nonumber\\&&{}
    -\re\left[ 1.3464\,\bu \D x{\bv } 
    +1.5041\,{\bv }\D y{\bv } +0.1577\,{\bv }\D x{\bu } 
    \right.\nonumber\\&&\left.\quad{}
    +0.1483\frac{\bv }{\eta}\left(\bu {\eta}_x+{\bv }{\eta}_y\right) 
    \right]
    \nonumber\\&&{}
    +\left[ \DD x{\bv } +4.0930\DD y{\bv } 
    +3.0930\frac{\partial^2\bu }{\partial x\partial y}  
    \right.\nonumber\\&&\left.\quad{}
    +4.8333\frac{\eta_ y}{\eta}\D y{\bv } 
    +\frac{\eta_x}{\eta}\D x{\bv } 
    +1.9167\frac{\eta_x}{\eta}\D y{\bu } 
    +1.9167\frac{h_ y}{\eta}\D x{\bu } 
    \right.\nonumber\\&&\left.\quad{}
    +\left( 
    -0.5033\frac{\eta_x^2}{\eta^2} 
    -\frac{\eta_{xx}}{2\eta} 
    +0.1061\frac{\eta_ y^2}{\eta^2} 
    -0.5834\frac{\eta_{y y}}{\eta} 
    \right)\bv  
    \right.\nonumber\\&&\left.\quad{}
    +\left( 0.6094\frac{\eta_ y\eta_x}{\eta^2} 
    -0.0833\frac{\eta_{x y}}{\eta} \right)\bu  \right]
    \,,
    \label{eq:flatv}
\end{eqnarray}
Observe that the substrate drag, gravitational and surface tension
terms are quite straightforward.  However, the self-advection terms
exhibit subtle interactions between the components of the velocity
fields due to the specific shape of the velocity profiles.  Further
subtleties occur in the viscous terms which not only show explicitly a
differential lateral dispersion of momentum, but also a complex
interaction with variations in the free surface shape.  

\begin{figure}[tbp]
    \centering
    \includegraphics[width=0.88\textwidth]{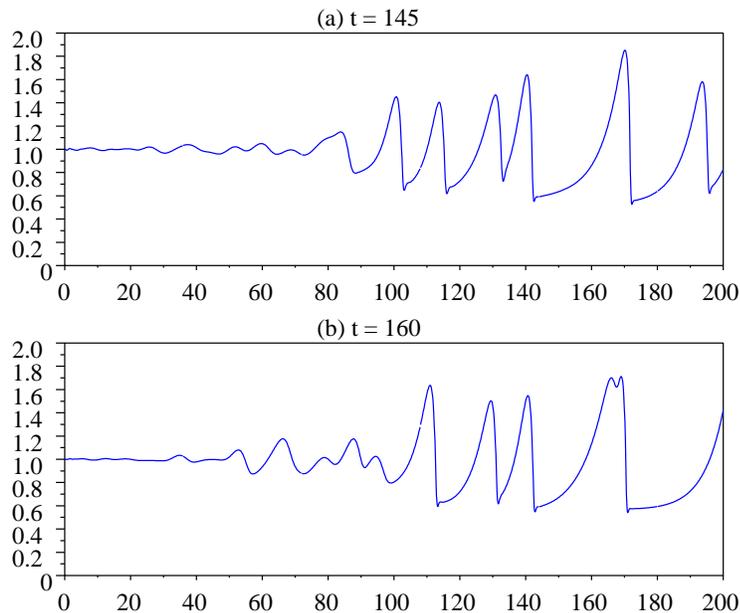}
	\caption{two instants of a 2D fluid falling down a vertical plane
	substrate with Reynolds number $\re=20$\,, gravity and Weber number
	$\gr=\we=3$\,.  The fluid thickness~$\eta$ as a function of
	distance~$x$ shows that white noise at the inlet $x=0$ is
	selectively amplified to solitary pulses that move and merge: the
	close pair of pulses near $x\approx130$ in~(a) at time $t=145$
	move and merge to the large pulse at $x\approx165$ in~(b) at time
	$t=160$\,.}
    \label{fig:twod}
\end{figure}
\paragraph{Wave transitions:} this model resolves 1D wave transitions
such as those reported by Chang, Demekhin and Saprikin~\cite{Chang02}.
Their parabolicised Navier-Stokes equation~(1) corresponds to our
non-dimensional Navier-Stokes equation~(\ref{eq:nonns}) with
$\gr=\we=3$ and our Reynolds number $\re=15\delta$ for Chang
et~al's~$\delta$.  See the corresponding simulations of our
model~(\ref{eq:flath}--\ref{eq:flatu}) restricted to 2D flow (as
in~\cite{Roberts94c}) and forced by white noise at the inlet, although
the forcing here is a little larger than that of Chang
\etal~\cite{Chang02}: the evolution and merger of the solitary pulses
are qualitatively the same as that simulated by Chang \etal\ The only
noticeable difference is that the retained dissipation in our modelling
almost entirely removes the surface tension waves in front of the
solitary pulses.

\begin{figure}[tbp]
    \centering
    \begin{tabular}{cc}
        \includegraphics[width=0.46\textwidth]{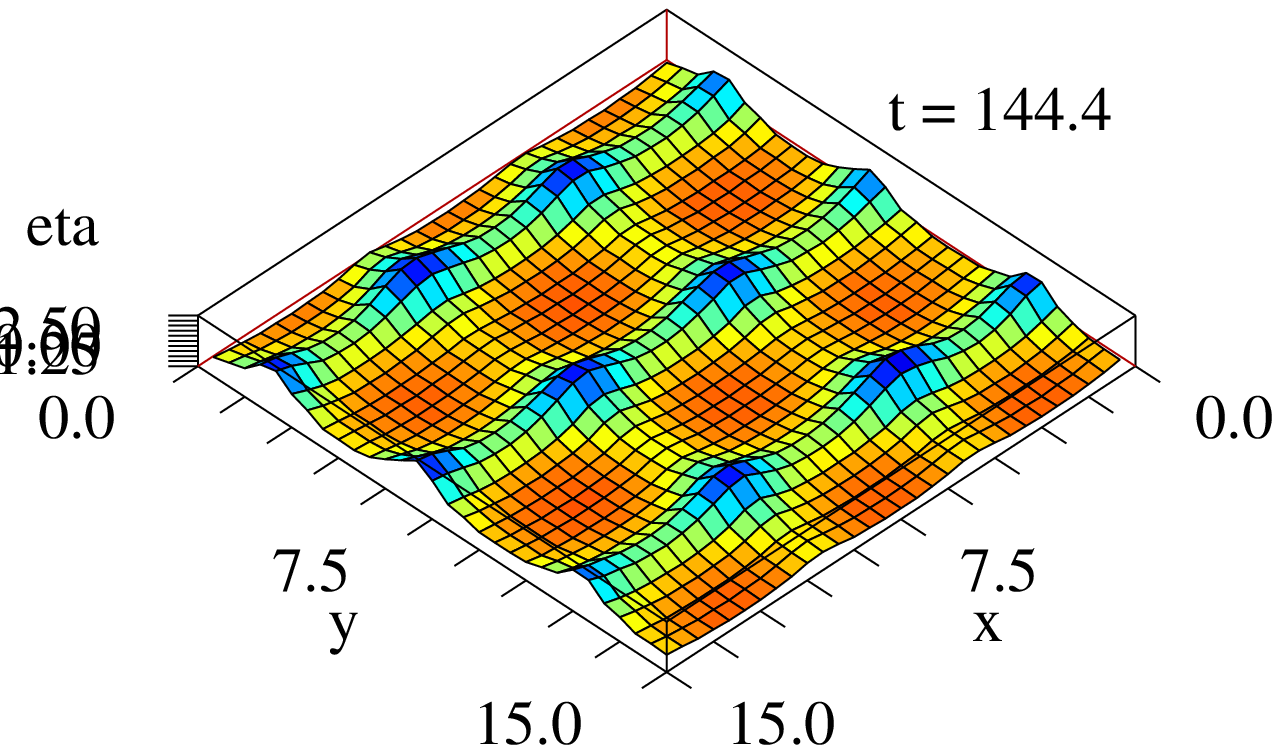} & 
        \includegraphics[width=0.46\textwidth]{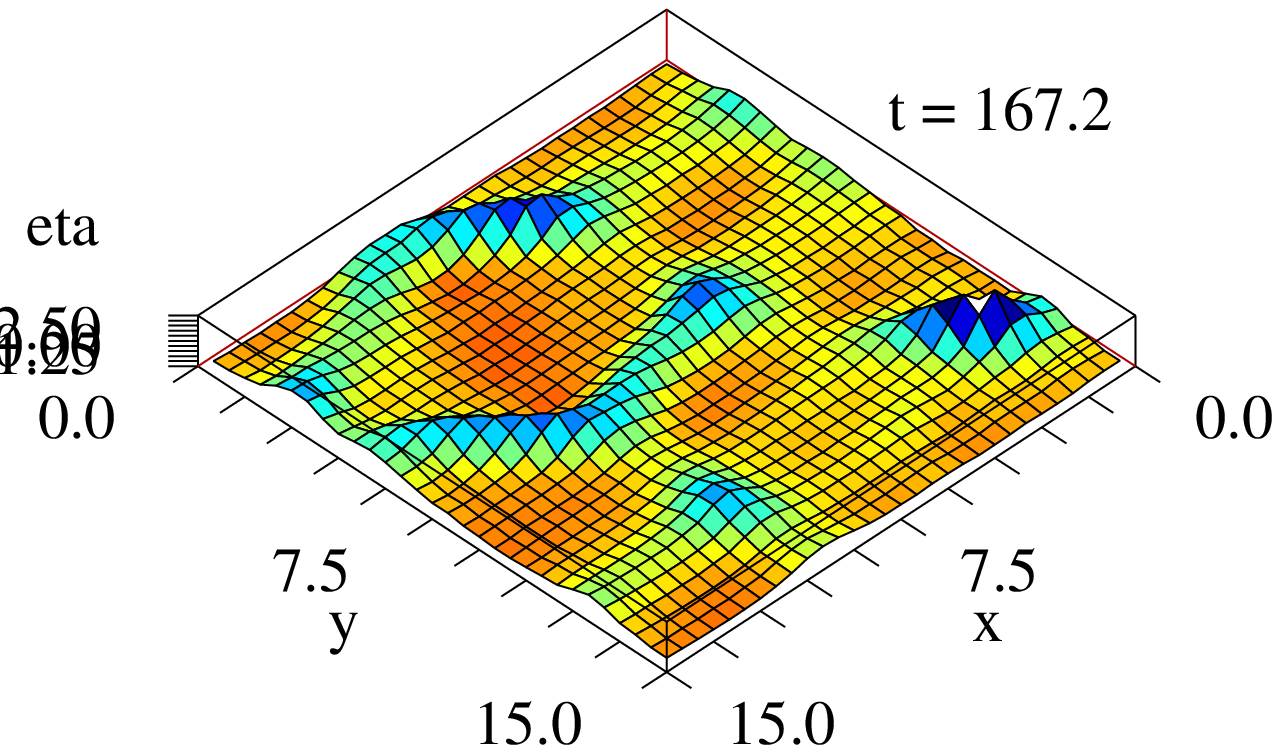}
    \end{tabular}
   \caption{two instants of unsteady waves on a vertically vibrating
   flat plate simulating the Faraday waves vacillating between
   ordered patterns (left) and irregular patterns (right): for mean
   fluid thickness~$1$, no surface tension, $\re=\gr=40$\,, and
   normal gravity modulated by the factor $1+0.55\,\sin(2.2\,t)$\,.}
    \label{fig:fara}
\end{figure}
\paragraph{Faraday waves} Vigorous vertical vibration of a layer of
fluid on a flat plate leads to a rich repertoire of spatio-temporal
dynamics~\cite[e.g.]{Miles90a,Perlin00}, such as those shown in
Figure~\ref{fig:fara}.  We choose the reference velocity to be the
shallow water wave speed $U=\sqrt{gH}$ then the non-dimensional
parameters $\re=\gr$\,.  Achieve the vertical vibration simply by
modulating the normal gravity in~(\ref{eq:flatu}--\ref{eq:flatv}) by,
for example, the factor $1+0.55\,\sin(2.2\,t)$\,.  This frequency is
roughly twice that of waves with wavelength~$5$ and see in
Figure~\ref{fig:fara} that these waves are generated by a Mathieu-like
instability.  But involved nonlinear interactions lead to complex
evolution of the spatial pattern of waves.

\paragraph{Vorticity and divergence} We explore some more of the
dynamics described by the model~(\ref{eq:flath}--\ref{eq:flatv}).
Consider the linearised dynamics of small variations on a film of
nearly constant thickness: $\eta=1+h(x,y,t)$ where $h$~and the lateral
velocity~$\bbu$ are small.  The linearised versions
of~(\ref{eq:flath}--\ref{eq:flatv}) are:
\begin{eqnarray*}
    h_t & = & -\bu_x-\bv_y\,,  \\
    \re\bu_{t} & = & -\frac{\pi^2}{4}\bu
    +\frac{\pi^2}{12}\left[ \gr (g_x+g_nh_x)
    +\we\left(h_{xxx}+h_{xyy}\right) \right]
    \\&&{}
    +(1+\varpi)\bu_{xx}+\bu_{yy}+\varpi\bv_{xy}\,,\\
    \re\bv_{t} & = & -\frac{\pi^2}{4}\bv
    +\frac{\pi^2}{12}\left[ \gr (g_y+g_nh_y)
    +\we\left(h_{xxy}+h_{yyy}\right) \right]
    \\&&{}
    +\bv_{xx}+(1+\varpi)\bv_{yy}+\varpi\bu_{xy}\,,
\end{eqnarray*}
where $\varpi=3.0930$ (the name $\varpi$ is chosen 
because its value is coincidentally close to~$\pi$).
\begin{itemize}
    \item
First take $\partial_y$ of the second from $\partial_x$ of the third to
deduce equation~(\ref{eq:vvort}) governing the mean normal
vorticity~$\bw=\bv_x-\bu_y$\,.  Observe that linearly it is decoupled
from the other components of the fluid dynamics: the mean normal
vorticity simply decays by drag on the substrate and by lateral
diffusion.

\item Conversely, the first of the linearised equations together with
the divergence of the second two equations decouple from the mean
normal vorticity to give~(\ref{eq:mcont}--\ref{eq:mdiv}) for the film
thickness and the mean flow divergence~$\bd=\bu_x+\bv_y$ as discussed
briefly in the Introduction.  A little analysis shows that in the
absence of gravity ($\gr=0$) this model predicts damped waves for
lateral wavenumbers
\begin{displaymath}
    a>a_c=\frac{\pi/2}{\sqrt{\pi\sqrt{\re\we/3}-(1+\varpi)}}\,.
\end{displaymath}
Numerical solutions of the physical eigenvalue problem agree closely
with this for $\re\we>30$ even though the critical wavenumber is as
large as $a_c\approx 0.65$\,.  Recall that the limit~(\ref{eq:logder})
on logarithmic derivatives in this model implies the wavenumber must be
less than~$1.9/\eta$; here the critical~$a_c\approx 0.65$ on a fluid of
depth near~$1$ is comfortably within the limit.  Waves cannot be
captured by the single mode of a lubrication model such as~(\ref{roy}).


\item Lastly, in this linear approximation, lateral components of 
gravity just induce a mean flow in the direction of the lateral 
component.
\end{itemize}
Substrate curvature and the nonlinear effects of advection and
large-scale variations in the thickness modify this description of the
basic dynamics of the fluid film.

\paragraph{Radial flow with axisymmetry:} turn on a tap producing a
steady stream into a basin with a flat bottom: you will see
the flow spreads out in a thin layer, then at some radial distance it
undergoes a hydraulic jump to a thicker layer spreading more
slowly~\cite[e.g.]{Watanabe00}. A model with inertia is essential for
modelling such a hydraulic jump.  Here use polar coordinates
$(r,\theta)$, whence the substrate has zero curvature $k_1=k_2=0$\,,
but scale factors~$m_1=1$ and~$m_2=r$\,.  Then describe axisymmetric
dynamics by neglecting angular flow and variations while retaining the
radial velocity~$\bar u$:
\begin{eqnarray*}
    \eta_t
    & = & -\frac{1}{r}{\partial_r}(r\bar u\eta)\,,  \\
    {\cal R}\bar u_t & = & -\frac{\pi^2}{4}\frac{\bar u}{\eta^2}
    +1.4167\frac{\eta_r}{r\eta}\bar u
    +\frac{\pi^2}{12}\left[ \gr g_n\eta_r 
    +\we\partial_r\left\{\frac{1}{r}\partial_r\left(r\eta_r\right)\right\} \right]
    \\&&{}
    -{\cal R}\left[ 1.5041\,\bar u\bar u_r +0.1577\frac{1}{r}{\bar 
    u}^2 +0.1483\frac{\eta_r}{\eta}{\bar u}^2 \right]
    \\&&{}
    +\left[
    4.0930\,\partial_r\left\{\frac{1}{r}\partial_r\left(r\bar 
    u\right)\right\} +4.8333\frac{\eta_r}{\eta}{\bar u}_r
    +\left( 0.1061\frac{\eta_r^2}{\eta^2} -0.5834\frac{\eta_{rr}}{\eta^2}
    \right)\bar u\right]\,.
\end{eqnarray*}
\begin{figure}[tbp]
    \centering
    \includegraphics{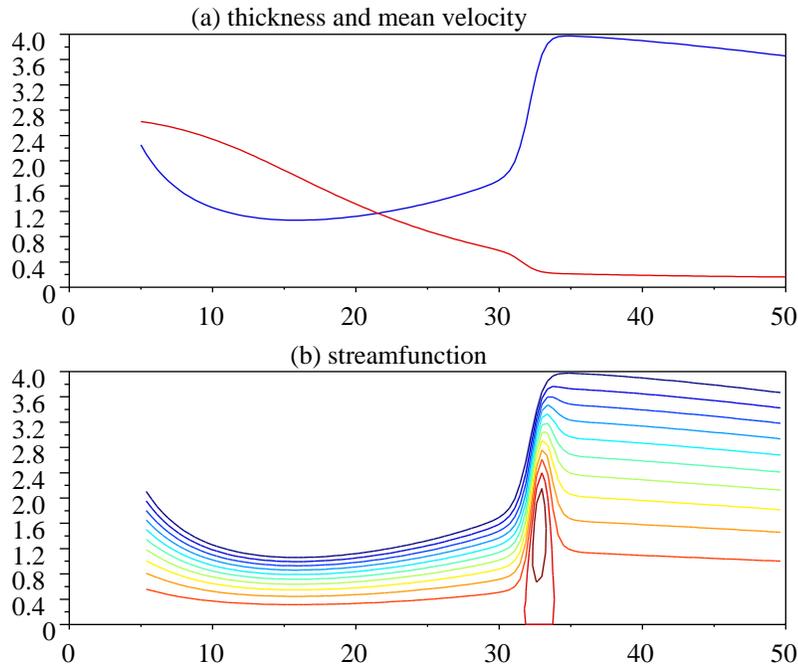}
	\caption{steady axisymmetric radial flow on a flat substrate:
	(a)~free surface thickness~$\eta$ and mean velocity~$\bar u$ versus
	radius~$r$; (b)~streamlines showing a recirculation under the
	hydraulic jump at $r\approx33$\,.  The Reynolds number ${\cal
	R}=15$\,, gravity number~$\gr=1$ and no surface tension,
	$\we=0$\,.}
    \label{fig:steady}
\end{figure}

The above equations may be integrated in time to see evolving dynamics
in a radial flow.  Here we focus only upon steady solutions obtained by
fixing an inlet condition of flow leaving the faucet with prescribed
thickness and velocity; in Figure~\ref{fig:steady} the flow has
$\eta=2.25$ and $\bar u=2.62$ at radius $r=5$\,, and exiting the domain
with some prescribed mean velocity at large distance, $\bar u=0.16$
at~$r=50$ in Figure~\ref{fig:steady}.  Newton iteration then finds the
solutions for fluid thickness~$\eta(r)$ and mean velocity~$\bar u(r)$
shown in Figure~\ref{fig:steady}(a).  See the flow spreads out in a
thin ($\eta\approx 1$) fast flow before undergoing a hydraulic jump at
distance~$r\approx 33$ to a thick ($\eta\approx 4$) slow flow.  The
streamlines in in Figure~\ref{fig:steady}(b), obtained from the
velocity fields~(\ref{uex}--\ref{vex}), show the presence of a
recirculation under the jump as also seen in the experiments cited by
Watanabe, Putkaradze and Bohr~\cite{Watanabe00}.  Our model expressed
in depth averaged quantities resolves such nontrivial internal flow
structures.

The steady flow in Figure~\ref{fig:steady} is near the limit of
applicability of the model.  Although the free surface looks steep in
the figure, the slope is everywhere less than~$1.08$ which, although
less than the limit~(\ref{eq:logder}) identified earlier, is about as
large as one could reasonably use.  For interest, other lateral
derivatives have the following ranges: $\eta_{rr}\in[-.88,.74]$\,,
$\bar u_r\in[-.15,0]$ and $\bar u_{rr}\in[-.08,.11]$\,.  We also find
the hydraulic jump with recirculation in less extreme flows than the
one shown in Figure~\ref{fig:steady}.  However, we have not yet found
any flows with an extra eddy at the surface of the hydraulic jump as
reported for experiments with larger jumps~\cite[\S2.1]{Watanabe00}.

\subsection{Flow outside a cylinder resolves evolving beads}
\label{Scyl}

Thin film flow on the outside or the inside of circular cylinders or 
tubes are important in a number of biological and engineering 
applications.  For example, Jensen~\cite{Jensen97} studied the effects 
of surface tension on a thin liquid layer lining the interior of a 
cylindrical and curving tube and derived a corresponding evolution 
equation in the lubrication approximation.  Our 
model~(\ref{ldm3d1}--\ref{ldm3d2}) could be used to extend his 
modelling to flows with inertia.  Similarly, Atherton \& 
Homsy~\cite{Atherton76}, Kalliadasis \& Chang~\cite{Kalliadasis94} 
and Kliakhandler \etal~\cite{Kliakhandler01}
considered coating flow down vertical fibres and similarly derived 
nonlinear lubrication models.  Here we record the 
model~(\ref{ldm3d1}--\ref{ldm3d2}) as it appears in full for flows 
both inside and outside a circular cylinder.  The specific model for a 
circular tube which is bent and twisted is left for later work.

Use a coordinate system with $s$~the axial coordinate and $\theta$~the
angular coordinate; denote the averaged axial and angular velocities
by~$\bbu$ with components by~$\bar u$ and~$\bv$, respectively.  The
substrate has scale factors~$m_1=1$ and~$m_2=R$ where~$R$ is the
radius of the cylinder, and curvatures~$k_1=0$ and~$k_2=\mp1/R$ where
the upper/lower sign is for flow outside/inside of the cylinder.  Then
the model on a cylinder is, where here $\zeta=\eta\pm\eta^2/(2R)$\,,
\begin{eqnarray}
    \D t\zeta & \approx & -\D s{(\eta\bu )}-\frac{1}{R}\D\theta{(\eta 
    \bv )}\,,
    \label{eq:cylh}  \\
    \re\D t{\bu } & \approx & 
    -\frac{\pi^2}{4}\frac{\bu }{\eta^2} 
    \pm\frac{\bu }{R\eta} -0.6487\frac{\bu }{R^2}
    +\we\frac{\pi^2}{12}\left[ \frac{1}{R^2}\eta_s 
    +\eta_{sss}+\frac{1}{R^2}\eta_{s\theta\theta} \right]
    \nonumber\\&&{}
    +\gr\left[\frac{\pi^2}{12}\left( g_s +g_n\eta_s \right)
    \pm0.4891\frac{g_s\eta}{R} \right]
    \nonumber\\&&{}
    -\re\left[ 1.5041\,\bu \D s{\bu } +1.3464\frac{\bar 
    v}{R}\D\theta{\bu } +0.1577\frac{\bu }{R}\D\theta{\bv } 
    \right.\nonumber\\&&\left.\quad{}
    +0.1483\frac{\bu }{\eta}\left( \bu \eta_s
    +\frac{\bv }{R}\eta_\theta \right) \right]
    \nonumber\\&&{}
    +\left[ 4.0930\DD s{\bu } +\frac{1}{R^2}\DD \theta{\bu } 
    +3.0930\frac{1}{R}\frac{\partial^2\bv }{\partial s\partial\theta}
    \right.\nonumber\\&&\left.\quad{}
    +4.8333\frac{\eta_s}{\eta}\D s{\bu } 
    +\frac{\eta_\theta}{R^2\eta}\D\theta{\bu } 
    +1.9167\frac{\eta_s}{R\eta}\D\theta{\bv } 
    +1.9167\frac{\eta_\theta}{R\eta}\D s{\bv } 
    \right.\nonumber\\&&\left.\quad{}
    +\left( 
    -0.5033\frac{\eta_\theta^2}{R^2\eta^2} 
    -\frac{\eta_{\theta\theta}}{2R^2\eta} 
    +0.1061\frac{\eta_s^2}{\eta^2} -0.5834\frac{\eta_{ss}}{\eta} 
    \right)\bu  
    \right.\nonumber\\&&\left.\quad{}
    +\left( 0.6094\frac{\eta_\theta\eta_s}{R\eta^2} 
    -0.0833\frac{\eta_{s\theta}}{R\eta} \right)\bv  \right]
    \,,
    \label{eq:cylu}  \\
    \re\D t{\bv } & \approx & 
    -\frac{\pi^2}{4}\frac{\bv }{\eta^2} 
    \pm\frac{3\bv }{R\eta} -2.8381\frac{\bv }{R^2}
    +\we\frac{\pi^2}{12}\left[ \frac{1}{R^2}\frac{\eta_\theta}{R} 
    +\frac{1}{R}\eta_{ss\theta} +\frac{1}{R^3}\eta_{\theta\theta\theta} \right]
    \nonumber\\&&{}
    +\gr\left[ \frac{\pi^2}{12}\left( g_\theta 
    +g_n\frac{\eta_\theta}{R} \right)
    \pm0.2337\frac{g_\theta\eta}{R} \right]
    \nonumber\\&&{}
    -\re\left[ 1.3464\,\bu \D s{\bv } +1.5041\frac{\bar 
    v}{R}\D\theta{\bv } +0.1577\,{\bv }\D s{\bu } 
    \right.\nonumber\\&&\left.\quad{}
    +0.1483\frac{\bv }{\eta}\left( \bu \eta_s
    +\frac{\bv }{R}\eta_\theta \right) \right]
    \nonumber\\&&{}
    +\left[ \DD s{\bv } +4.0930\frac{1}{R^2}\DD\theta{\bv } 
    +3.0930\frac{1}{R}\frac{\partial^2\bu }{\partial s\partial\theta}  
    \right.\nonumber\\&&\left.\quad{}
    +4.8333\frac{\eta_\theta}{R^2\eta}\D \theta{\bv} 
    +\frac{\eta_s}{\eta}\D s{\bv } 
    +1.9167\frac{\eta_s}{R\eta}\D\theta{\bu } 
    +1.9167\frac{\eta_\theta}{R\eta}\D s{\bu } 
    \right.\nonumber\\&&\left.\quad{}
    +\left( 
    -0.5033\frac{\eta_s^2}{\eta^2} 
    -\frac{\eta_{ss}}{2\eta} 
    +0.1061\frac{\eta_\theta^2}{R^2\eta^2} 
    -0.5834\frac{\eta_{\theta\theta}}{R^2\eta} 
    \right)\bv  
    \right.\nonumber\\&&\left.\quad{}
    +\left( 0.6094\frac{\eta_\theta\eta_s}{R\eta^2} 
    -0.0833\frac{\eta_{s\theta}}{R\eta} \right)\bu  \right]
    \,,
    \label{eq:cylv}
\end{eqnarray}
For a nontrivial example, see the beading of fluid on a thin fibre in
Figures~\ref{fig:cyl4} and~\ref{fig:cyl8}: gravity first rather quickly
moves a lump of fluid to below the horizontal cylinder of the fibre;
thereafter, surface tension more slowly gathers more fluid into the
beading fluid which beads both below and above the fibre; gravity
causes the fluid bead to stabilise off the centre of the fibre
(curiously, there is more in the bead above the fibre than in the
original lump); and finally the bead slides along the fibre as it is
angled downwards a little.  The three time scales in this evolution,
the fast gravity forced flow, the slower surface tension driven flow,
and the even longer term sliding, are captured in this model.

Simply obtain axisymmetric flows by setting to zero any 
derivatives with respect to~$\theta$, and also setting 
$g_n=g_{\theta}=0$ as non-zero values would break the symmetry.
The equation for~$\bv $ then just describes the decay of angular 
flow around the cylinder so also set~$\bv =0$\,.
Thus the axisymmetric model is 
\begin{eqnarray}
    \D t\zeta & \approx & -\D s{(\eta\bu )}\,,
    \label{eq:axih}  \\
    \re\D t{\bu } & \approx & 
    -\frac{\pi^2}{4}\frac{\bu }{\eta^2} 
    \pm\frac{\bu }{R\eta} -0.6487\frac{\bu }{R^2}
    +\we\frac{\pi^2}{12}\left[ \frac{1}{R^2}\eta_s +\eta_{sss} \right]
    \nonumber\\&&{}
    +\gr\left[ \frac{\pi^2}{12}g_s \pm0.4891\frac{g_s\eta}{R} \right]
    \nonumber\\&&{}
    -\re\left[ 1.5041\,\bu \D s{\bu }  
    +0.1483\frac{\bu }{\eta}\bar{u}\D s\eta \right]
    +\left[ 4.0930\DD s{\bu } 
    \right.\nonumber\\&&\left.\quad{}
    +4.8333\frac{\eta_s}{\eta}\D s{\bu } 
    +\left( 
    0.1061\frac{\eta_s^2}{\eta^2} -0.5834\frac{\eta_{ss}}{\eta} 
    \right)\bu   \right]
    \,;
    \label{eq:axiu}  
\end{eqnarray}
recall that the upper/lower sign is for flow outside/inside of the
cylinder.  As in lubrication models~\cite[p254]{Roy96}, see that
surface tension on the cylinder acts through the term $\we \eta_s/R^2$
in~(\ref{eq:axiu}) or~(\ref{eq:cylu}) rather like a radially outwards
body force such as the term $\gr g_n\eta_s$ in~(\ref{eq:cylu}).

\subsection{Flow about a small channel grows vortices}
\label{sec:chan}

Consider the flow on a substrate with a small channel aligned downhill.
We compare this viscous flow with the high Reynolds number
experiments of Bousmar~\cite{Bousmar02, Bousmar03} who modelled
turbulent flow over flood plains and channels in a flume with water of
variable depth but of the order of 5\,cm~deep.

\begin{figure}[tbp]
    \centering
\begin{tabular}{cc}
        (a)&\includegraphics[height=0.4\textheight]{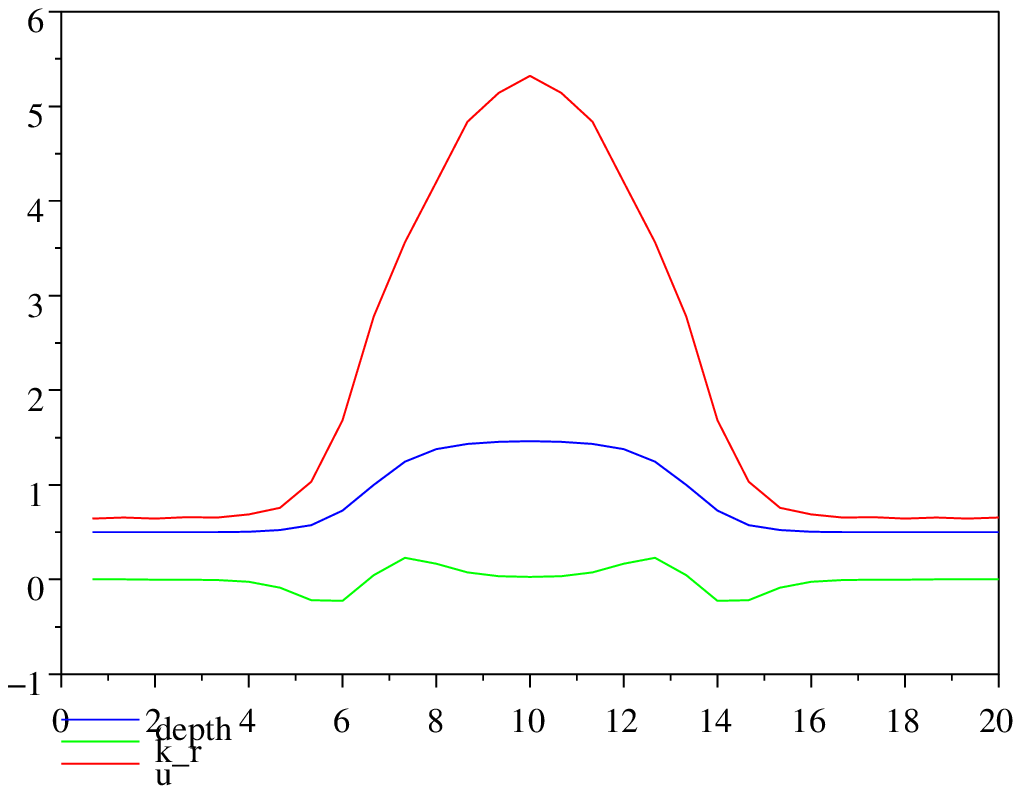} \\
        (b)&\includegraphics[height=0.4\textheight]{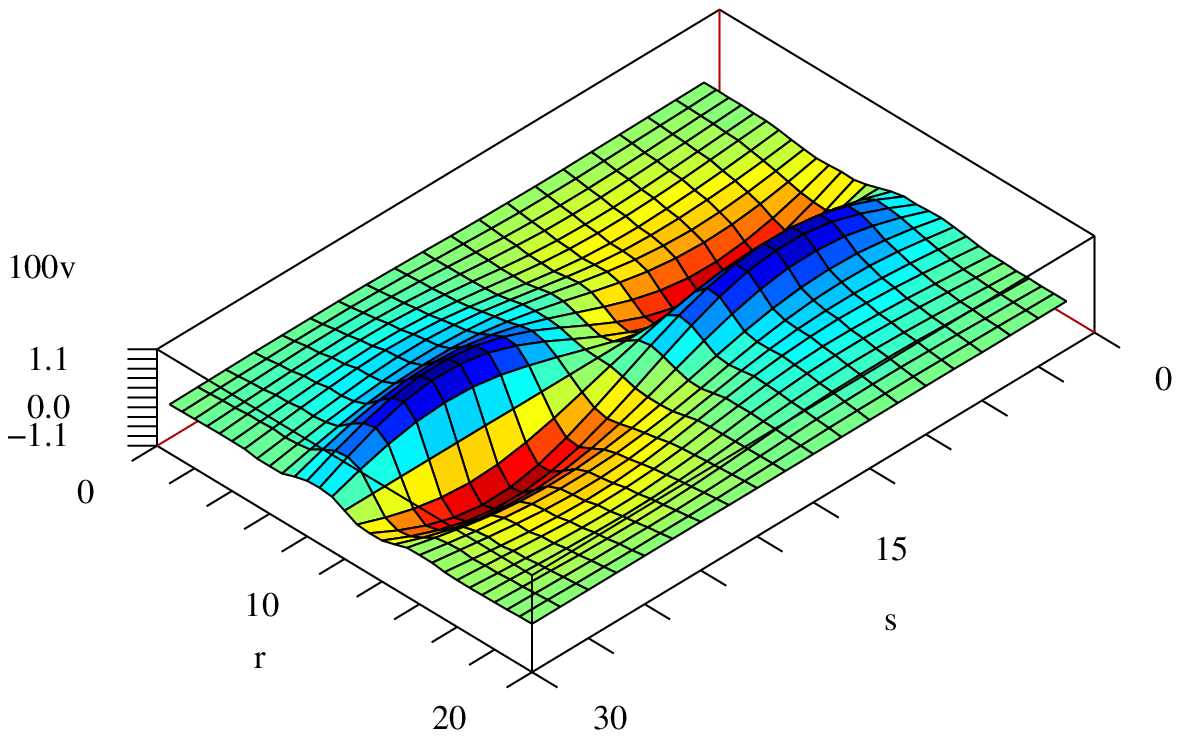}
\end{tabular}
	\caption{(a) an example channel is three times as deep in the
	middle as the surrounding shallows, blue; the substrate has
	curvature~$k_r$, green; and the fluid flow is nearly ten times as
	fast in the channel as in the shallows, red; (b)~this base flow is
	unstable to superimposed travelling vortices on the shear near the
	sides of the channel as shown by the cross-channel velocity.  Here
	$\re=\gr=80$ on a substrate sloping down at
	angle~$\vartheta=0.1$~radians, and with no surface tension,
	$\we=0$\,.}
    \label{fig:chan}
\end{figure}

First create the coordinate system.  Bousmar's channel and flood plain
had constant shape along the stream, the depth only varied across the
stream.  Thus here let $s=x_1$ be the along stream coordinate, $r=x_2$
be the horizontal distance across the stream on the
substrate,\footnote{There is no good reason for using the variable
name~$r$ for distance horizontally across the stream, only that it is
next to the letter~$s$ in the alphabet.  In this subsection the
variable~$r$ is \emph{not} used to indicate any sort of radius.} and
$y$~measure distance normal to the substrate.  The curved substrate is
located a distance~$d(r)>0$ from the $sr$-plane in a normal direction,
see the example $d(r)$ in the middle curve of Figure~\ref{fig:chan}(a).
Using, for this subsection, $\vec j$, $\vec k$ and~$\vec i$ as the
vertical and two horizontal unit vectors, the unit vectors, scale
factors and curvature of the substrate coordinate system
are\footnote{Useful relationships are: $m_r^2=1+{d'}^2$\,,
$d''=-k_rm_r^3$\,, and $m_r'=-d'k_rm_r^2$\,.}
\begin{eqnarray*}
    \vec e_s=\vec k\,,
    &&\vec e_r=\frac1{\sqrt{1+{d'}^2}}(\vec i-d'\vec j)\,,
    \qquad \vec e_n=\frac1{\sqrt{1+{d'}^2}}(d'\vec i+\vec j)\,,
    \\
    m_s=1\,,
    &&m_r=\sqrt{1+{d'}^2}\,,
    \\
    k_s=0\,,
    &&k_r=-\frac{d''}{(1+{d'}^2)^{3/2}}\,.
\end{eqnarray*}
This expression for the curvature~$k_r$ is well known.  Note: the
normal coordinate~$y$ is \emph{not} the vertical coordinate, and so a
flat fluid surface located at, say, the location of the reference
$sr$-plane is represented by the varying $y=d({1+{d'}^2})^{1/2}$\,.
Similarly nontrivial, as the channel slopes down at an
angle~$\vartheta=0.1$~radians to the horizontal and not sideways
tilted, the gravitational forcing is in the direction
\begin{equation}
    \hat{\vec g}=\sin\vartheta\,\vec e_s
    +\frac{d'}{m_r}\cos\vartheta\,\vec e_r
    -\frac1{m_r}\cos\vartheta\,\vec e_n\,.
    \label{eq:chang}
\end{equation}
This coordinate system suits any flow where the substrate is almost
arbitrarily curved in only one direction, not just flow along a
channel.\footnote{The model~(\ref{eq:chah}--\ref{eq:chav}) reduces to
that for the flow outside/inside a cylinder,
(\ref{eq:cylh}--\ref{eq:cylv}), when $r=\theta$ and the substrate scale
factors are set to $k_r=\mp1/R$ and $m_r=R$ (only the direction of
gravity~(\ref{eq:chang}) is incorrect).  This algebraic connection
occurs despite the cylinder not  being strictly encompassed by a
depth~$d(r)$ below any reference plane.  Indeed the beading flow on a
cylinder shown in Figures~\ref{fig:cyl4} and~\ref{fig:cyl8} was
actually obtained using code for the
model~(\ref{eq:chah}--\ref{eq:chav}) of this subsection.}

Second, computer algebra gives the model on this substrate as, where
here $\zeta=\eta-k_r\eta^2/2$\,,
\begin{eqnarray}
    \D t\zeta & \approx & -\D s{(\eta\bu )}-\frac{1}{m_r}\D r{(\eta 
    \bv )}\,,
    \label{eq:chah}  \\
    \re\D t{\bu } & \approx & 
    -\frac{\pi^2}{4}\frac{\bu }{\eta^2} 
    -k_r\frac{\bu }{\eta} -0.6487\,k_r^2{\bu }
    \nonumber\\&&{}
    +\we\frac{\pi^2}{12}\left[ k_r^2\eta_s 
    +\eta_{sss}
    +\frac1{m_r}\D r{}\left(\frac1{m_r}\D r{\eta_s}\right)
    \right]
    \nonumber\\&&{}
    +\gr\left[\frac{\pi^2}{12}\left( g_s +g_n\eta_s \right)
    -0.4891\,k_r{g_s\eta} \right]
    \nonumber\\&&{}
    -\re\left[ 1.5041\,\bu \D s{\bu } +1.3464\frac{\bar 
    v}{m_r}\D r{\bu } +0.1577\frac{\bu }{m_r}\D r{\bv } 
    \right.\nonumber\\&&\left.\quad{}
    +0.1483\frac{\bu }{\eta}\left( \bu \eta_s
    +\frac{\bv }{m_r}\eta_r \right) \right]
    \nonumber\\&&{}
    +\left[ 4.0930\DD s{\bu } 
    +\frac{1}{m_r}\D r{}\left( \frac1{m_r}\D r{\bu} \right)
    +3.0930\frac{1}{m_r}\frac{\partial^2\bv }{\partial s\partial r}
    \right.\nonumber\\&&\left.\quad{}
    +4.8333\frac{\eta_s}{\eta}\D s{\bu } 
    +\frac{\eta_r}{m_r^2\eta}\D r{\bu } 
    +1.9167\frac{\eta_s}{m_r\eta}\D r{\bv } 
    +1.9167\frac{\eta_r}{m_r\eta}\D s{\bv } 
    \right.\nonumber\\&&\left.\quad{}
    +\left( 
    -0.5033\frac{\eta_r^2}{m_r^2\eta^2} 
    -\frac1{2\eta m_r}\D r{}\left( \frac{\eta_r}{m_r} \right)
    \right.\right.\nonumber\\&&\left.\left.\qquad{}
    +0.1061\frac{\eta_s^2}{\eta^2} 
    -0.5834\frac{\eta_{ss}}{\eta} 
    \right)\bu  
    \right.\nonumber\\&&\left.\quad{}
    +\left( 0.6094\frac{\eta_r\eta_s}{m_r\eta^2} 
    -0.0833\frac{\eta_{sr}}{m_r\eta} \right)\bv  \right]
    \,,
    \label{eq:chau}  \\
    \re\D t{\bv } & \approx & 
    -\frac{\pi^2}{4}\frac{\bv }{\eta^2} 
    -k_r\frac{3\bv }{\eta} -2.8381\,k_r^2{\bv }
    +\we\frac{\pi^2}{12}\left[ \frac{k_r'}{m_r}
    +k_r^2\frac{\eta_r}{m_r} 
    \right.\nonumber\\&&\left.\quad{}
    +\frac{\eta_{ssr}}{m_r} +\frac{1}{m_r}\D r{}\left(
    \frac1{m_r}\D r{}\left\{ \frac{\eta_r}{m_r} \right\} \right) 
    +2.7159\,\frac{\eta k_rk_r'}{m_r} \right]
    \nonumber\\&&{}
    +\gr\left[ \frac{\pi^2}{12}\left( g_r 
    +g_n\frac{\eta_r}{m_r} \right)
    -0.2337\,k_r{g_r\eta} \right]
    \nonumber\\&&{}
    -\re\left[ 1.3464\,\bu \D s{\bv } +1.5041\frac{\bar 
    v}{m_r}\D r{\bv } +0.1577\,{\bv }\D s{\bu } 
    \right.\nonumber\\&&\left.\quad{}
    +0.1483\frac{\bv }{\eta}\left( \bu \eta_s
    +\frac{\bv }{m_r}\eta_r \right) \right]
    \nonumber\\&&{}
    +\left[ \DD s{\bv } 
    +4.0930\frac{1}{m_r}\D r{}\left(\frac1{m_r}\D r{\bv}\right) 
    +3.0930\frac{1}{m_r}\frac{\partial^2\bu }{\partial s\partial r}  
    \right.\nonumber\\&&\left.\quad{}
    +4.8333\frac{\eta_r}{m_r^2\eta}\D r{\bv} 
    +\frac{\eta_s}{\eta}\D s{\bv } 
    +1.9167\frac{\eta_s}{m_r\eta}\D r{\bu } 
    +1.9167\frac{\eta_r}{m_r\eta}\D s{\bu } 
    \right.\nonumber\\&&\left.\quad{}
    +\left( 
    -0.5033\frac{\eta_s^2}{\eta^2} 
    -\frac{\eta_{ss}}{2\eta} 
    +0.1061\frac{\eta_r^2}{m_r^2\eta^2} 
    -0.5834\frac1{\eta m_r}\D r{}\left( \frac{\eta_r}{m_r} \right) 
    \right)\bv  
    \right.\nonumber\\&&\left.\quad{}
    +\left( 0.6094\frac{\eta_r\eta_s}{m_r\eta^2} 
    -0.0833\frac{\eta_{sr}}{m_r\eta} \right)\bu  \right]
    \,,
    \label{eq:chav}
\end{eqnarray}

Lastly, as expected, simulations show that fast flow develops in the
deeper channel and slow flow on the shallow regions, see
Figure~\ref{fig:chan}(a).  However, the shear in the mean downstream
velocity, Figure~\ref{fig:chan}(a), is unstable to relatively weak
horizontal vortices that grow in the shear and travel downstream, see
them in the mean lateral velocity shown in
Figure~\ref{fig:chan}(b);\footnote{The vortices apparent in
Figure~\ref{fig:chan}(b) fill the computational domain and thus may be
an artifice of the domain size.  However, otherwise identical
simulations on twice the channel length show twice as many vortices,
whereas simulations in a domain half as long again show a rich
modulation among vortices of roughly the shown length.  We deduce that
the displayed vortices are not solely an artifice of the computational
domain.} analogous notable vortices were observed by
Bousmar~\cite{Bousmar02, Bousmar03} in their turbulent flows.  As also
noted by Bousmar, see that the vortices here similarly extend into the
shallows, albeit weakly.

The simulation reported here has a change in depth of the substrate
sufficiently big so that the nonlinear nature of the derived model is
certainly essential: Decr\'e \& Baret~\cite[p155]{Decre03} comment that
nonlinear theories are needed in viscous flow if the change in
substrate profile is bigger than half the shallow fluid depth; here the
factor is about three, that is, six times the linear limit identified by
Decr\'e \& Baret.

\subsection{Flow on the outside of a sphere}
\label{Ssphere}

For flow on the outside of a sphere we use a coordinate system with
$\theta$ the co-latitude coordinate, $\phi$ the azimuthal (longitude)
coordinate, and co-latitude and azimuthal velocity components~$\bu$
and~$\bv $ respectively.  The substrate has scale factors~$m_1=R$
and~$m_2=R\sin\theta$ where~$R$ is the radius of the sphere, and
curvatures~$k_1=k_2=-1/R$\,.  Note: on a sphere every point is an
umbilical point; nonetheless, the earlier analysis is valid in this
conventional spherical coordinate system.  Then the model on a sphere
is, where here $\zeta=\eta+\eta^2/R+\eta^3/(3R^2)$\,,
\begin{eqnarray}
    \D t\zeta & \approx & -\frac{1}{R}\D \theta {(\eta\bu )}
    -\frac{1}{R\sin\theta}\D\phi{(\cos\theta\,\eta \bv )}\,,
    \label{eq:sphh}  \\
    \re\D t{\bu } & \approx & 
    -\frac{\pi^2}{4}\frac{\bu }{\eta^2} 
    -6.4718\frac{\bu }{R^2}
    +\gr\left[\frac{\pi^2}{12}\left( g_\theta  
    +\frac{g_n}{R} \eta_\theta  \right)
    +0.7228\frac{g_\theta \eta}{R}    \right]
    \nonumber\\&&{}
    +\we\frac{\pi^2}{12}\left[   
    -\frac{\cos2\theta}{R^3\sin^2\theta}\eta_\theta  
    +\frac{1}{R^3\sin\theta}\D\theta{}\left(\sin\theta\,\eta_{\theta\theta} \right)
    \right.\nonumber\\&&\left.\quad{}
    +\D\theta{}\left(\frac{1}{R^3\sin^2\theta}\eta_{\phi\phi}\right)
    \right]
    \nonumber\\&&{}
    -\re\left[ 1.5041\frac{\bu }{R}\D \theta {\bu } 
    +1.3464\left( \frac{\bv }{R\sin\theta}\D\phi{\bu }
    -\frac{\cos\theta}{R\sin\theta}{\bv }^2 \right)
    \right.\nonumber\\&&\left.\quad{}
    +0.1577\left( \frac{\bu }{R\sin\theta}\D\phi{\bv } 
    +\frac{\cos\theta}{R\sin\theta}{\bu }^2 \right)
    \right.\nonumber\\&&\left.\quad{}
    +0.1483\frac{\bu }{\eta}\left(\frac{\bu }{R}{\eta}_\theta 
    +\frac{\bv }{R\sin\theta}{\eta}_\phi\right) 
    \right]
    \nonumber\\&&{}
    +\left[ 4.0930\frac{1}{R^2}\left( \DD \theta {\bu } 
    +\frac{\cos\theta}{\sin\theta}\D\theta{\bu } 
    -\frac{\cos^2\theta}{\sin^2\theta}{\bu } \right)
    +\frac{1}{R^2\sin^2\theta}\DD \phi{\bu } 
    \right.\nonumber\\&&\left.\quad{}
    +3.0930\frac{1}{R^2\sin\theta}\frac{\partial^2\bv }{\partial \theta \partial\phi}
    -5.0930\frac{\cos\theta}{R^2\sin^2\theta}\D\phi{\bv}
    +4.8333\frac{\eta_\theta }{R^2\eta}\D \theta {\bu } 
    \right.\nonumber\\&&\left.\quad{}
    +\frac{\eta_\phi}{R^2\eta}\D\phi{\bu } 
    +\frac{\eta_\phi}{R^2\eta\sin^2\theta}\D\phi{\bu }
    +1.9167\frac{\eta_\theta }{R^2\sin\theta\,\eta}\D\phi{\bv } 
    \right.\nonumber\\&&\left.\quad{}
    +1.9167\frac{\eta_\phi}{R^2\sin\theta\,\eta}\D \theta {\bv } 
    \right.\nonumber\\&&\left.\quad{}
    +\left( 
    -0.5033\frac{\eta_\phi^2}{R^2\sin^2\theta\,\eta^2} 
    -\frac{\eta_{\phi\phi}}{2R^2\sin^2\theta\,\eta} 
    +0.1061\frac{\eta_\theta ^2}{R^2\eta^2} 
    \right.\right.\nonumber\\&&\left.\left.\qquad{}
    -0.5834\frac{\eta_{\theta \theta }}{R^2\eta}
    +1.4167\frac{\cos\theta}{R^2\sin\theta\,\eta}\eta_\theta
    \right)\bu  
    \right.\nonumber\\&&\left.\quad{}
    +\left( 0.6094\frac{\eta_\phi\eta_\theta }{R^2\sin\theta\,\eta^2} 
    -0.0833\frac{\eta_{\theta \phi}}{R^2\sin\theta\,\eta} 
    \right.\right.\nonumber\\&&\left.\left.\qquad{}
    -0.9167\frac{\cos\theta}{R^2\sin^2\theta\,\eta}\eta_\phi 
    \right)\bv  \right]
    \,,
    \label{eq:sphu}  \\
    \re\D t{\bv } & \approx & 
    -\frac{\pi^2}{4}\frac{\bv }{\eta^2} 
    +\frac{4\bv }{R\eta} 
    -3.3788\frac{\bv }{R^2}
    -\frac{\cos^2\theta}{R^2\sin^2\theta}{\bv }
    \nonumber\\&&{}
    +\gr\left[\frac{\pi^2}{12}\left( g_\phi 
    +{g_n}\frac{\eta_\phi}{R\sin\theta} \right)
    +0.7228\frac{g_\phi\eta}{R} \right]
    \nonumber\\&&{}
    +\we\frac{\pi^2}{12}\left[ \frac{2}{R^2}\frac{\eta_\phi}{R\sin\theta} 
    +\frac{1}{R^3\sin^2\theta}\D\theta{}\left( \sin\theta\,\eta_{\theta\phi} \right)
    +\frac{1}{R^3\sin^3\theta}\eta_{\phi\phi\phi} \right]
    \nonumber\\&&{}
    -\re\left[ 1.3464\frac{\bu }{R}\D \theta {\bv } 
    +1.5041\left( \frac{1}{R\sin\theta}\D\phi{\bv } 
    +\frac{\cos\theta}{R\sin\theta}{\bu } \right)\bv 
    \right.\nonumber\\&&\left.\quad{}
    +0.1577\,{\bv }\D \theta {\bu } 
    +0.1483\frac{\bv }{\eta}\left( \frac{\bu }{R}{\eta}_\theta 
    +\frac{\bv }{R\sin\theta}{\eta}_\phi \right) 
    \right]
    \nonumber\\&&{}
    +\left[ \frac{1}{R^2}\DD \theta {\bv } 
    +\frac{\cos\theta}{R^2\sin\theta}\D\theta{\bv }
    +4.0930\frac{1}{R^2\sin^2\theta}\DD\phi{\bv } 
    \right.\nonumber\\&&\left.\quad{}
    +5.0930\frac{\cos\theta}{R^2\sin^2\theta}\D\phi{\bu }
    +3.0930\frac{1}{R^2\sin\theta}\frac{\partial^2\bu }{\partial \theta \partial\phi}  
    \right.\nonumber\\&&\left.\quad{}
    +4.8333\frac{\eta_\phi}{R^2\sin^2\theta\,\eta}\D \phi{\bv} 
    +\frac{\eta_\theta }{R^2\eta}\D \theta {\bv } 
    +1.9167\frac{\eta_\theta }{R^2\sin\theta\,\eta}\D\phi{\bu } 
    \right.\nonumber\\&&\left.\quad{}
    +1.9167\frac{\eta_\phi}{R^2\sin\theta\,\eta}\D \theta {\bu }
    \right.\nonumber\\&&\left.\quad{}
    +\left( 
    -0.5033\frac{\eta_\theta ^2}{R^2\eta^2} 
    -\frac{\eta_{\theta \theta }}{2R^2\eta} 
    +0.1061\frac{\eta_\phi^2}{R^2\sin^2\theta\,\eta^2} 
    \right.\right.\nonumber\\&&\left.\left.\qquad{}
    -0.5834\frac{\eta_{\phi\phi}}{R^2\sin^2\theta\,\eta} 
    -\frac{5\cos\theta}{2R^2\sin\theta\,\eta}\eta_\theta
    \right)\bv  
    \right.\nonumber\\&&\left.\quad{}
    +\left( 0.6094\frac{\eta_\phi\eta_\theta }{R^2\sin\theta\,\eta^2} 
    -0.0833\frac{\eta_{\theta \phi}}{R^2\sin\theta\,\eta} 
    \right.\right.\nonumber\\&&\left.\left.\qquad{}
    +4.9167\frac{\cos\theta}{R^2\sin^2\theta\,\eta}\eta_\phi
    \right)\bu  \right]
    \,,
    \label{eq:sphv}
\end{eqnarray}
These models look horribly complicated but recall that depending upon
the application, simpler truncations are often appropriate; two such
examples are~(\ref{my2}) and~(\ref{eq:utrunc}).  These models have the
assurance of centre manifold theory that all physical effects are
included to the controlable specified accuracy.

\section{Conclusion}

We systematically analysed the Navier-Stokes equations for the flow of
a thin layer of a Newtonian fluid over an arbitrarily curved substrate.
The resulting general model~(\ref{ldm3d1}--\ref{ldm3d2}) resolves the
dynamical effects and interactions of inertia, surface tension, and a
gravitational body force as well as the substrate curvature.  We
presented evidence towards the end of \S\ref{ldm3d} that this model
applies to flows where the lateral gradients of the fluid thickness are
somewhat less than~$2$, see the more precise limit~(\ref{eq:logder}),
and (in \S\ref{cmam}) where the time scales of the flow are reasonably
longer than the decay of the second lateral shear mode, that is, longer
than $0.045\,\eta^2/\nu$\,.  The centre manifold paradigm for dynamical
modelling is based upon actual solutions of the governing Navier-Stokes
equations, parametrised in terms of cross-layer averages.  Further the
paradigm implicitly arranges the interaction terms between various
physical processes to support flexible truncation of the model as
appropriate for different parameter regimes; thus the relatively
complex model~(\ref{ldm3d1}--\ref{ldm3d2}) may be justifiably
simplified as needed by your application.

To illustrate a range of applications we briefly reported some
simulations of: wave transitions on a sloping substrate, Faraday waves
on a vibrating flat plate, and a viscous hydraulic jump in radial flow,
see~\S\ref{Sflat}; the formation and sliding of beads on a cylindrical
fibre with surface tension and gravity, see~\S\ref{Scyl}; and the
generation of vortices in the shear flow between a channel and
surrounding shallows, see~\S\ref{sec:chan}.  These simulations
demonstrate the resolution of the complex interactions between the
varied physical processes encompassed by the model.

\paragraph{Acknowledgment:} we thank the Australian Research Council 
for a grant to help support this work.

\appendix

\section{Computer algebra derives the model}
\label{3dp}

The \textsc{reduce}\footnote{At the time of writing, information about
\textsc{reduce} was available from Anthony C.~Hearn, RAND, Santa
Monica, CA~90407-2138, USA. \protect\url{mailto:reduce@rand.org}}
program which performs the derivation of centre manifold model is
listed below.  A little explanation is usefully given first.  Observe
that program variables are typeset in \verb|teletype| font.
\begin{itemize}
    \item The physical coordinate system within the program is 
    $(x_1,x_2,y)=(\verb|x|,\verb|z|,\verb|y|)$ with scale factors 
    $(h_1,h_2,h_3)=(\verb|h1|,\verb|h2|,1)$\,, velocity field 
    $(u_1,u_2,v)=(\verb|u|,\verb|w|,\verb|v|)$ and pressure 
    $p=\verb|p|$\,.  Scale factors of the substrate are $m_1=\verb|m1|$ and 
    $m_2=\verb|m2|$\,, whereas those evaluated on the free surface are 
    $\tilde h_1=\verb|hh1|$ and~$\tilde h_2=\verb|hh2|$\,.  
    
    Substrates with specific geometries, as discussed in 
    \S\ref{Sspec}, may be derived simply by coding information 
    about their curvatures~$k_i$ and scale factors~$m_i$ as shown for 
    the three cases discussed in \S\ref{Sspec}.

    \item Expressions are written in terms of a stretched coordinate 
    system~$X=x_1$, $Z=x_2$, $Y=y/\eta$, $T=t$ so that the free 
    surface of the fluid film is simply~$Y=1$\,.  In the program we use 
    $(X,Z,Y,T)=(\verb|xs|,\verb|zs|,\verb|ys|,\verb|ts|)$.  Then 
    spatio-temporal derivatives transform by the chain rule
    \begin{displaymath}
        \D{x_1}{}=\D X{}-\frac{Y\eta_X}{\eta}\D Y{}\,,\quad
	\D{x_2}{}=\D Z{}-\frac{Y\eta_Z}{\eta}\D Y{}\,,
    \end{displaymath}
    \begin{displaymath}
        \D{t}{}=\D T{}-\frac{Y\eta_T}{\eta}\D Y{}\,,\quad
	\mbox{and}\quad
	\D{y}{}=\frac{1}{\eta}\D Y{}\,,
    \end{displaymath}
    as coded.

    \item The amplitudes of the model are $(\eta,\bu _1,\bar 
    u_2)=(\verb|h|,\verb|uu|,\verb|ww|)$ with \verb|h(m,n)| denoting 
    the spatial derivative $\frac{\partial^{m+n} \eta}{\partial X^m 
    \partial Z^n}$ and similarly for \verb|uu(m,n)| and 
    \verb|ww(m,n)|.  The evolution of these quantities is given by
        \begin{displaymath}
        \D t\eta=\verb|gh|\,,\quad
	\D t{\bu _1}=\verb|gu|\,,\quad\mbox{and}\quad
	\D t{\bu _2}=\verb|gw|\,.
    \end{displaymath}
\end{itemize}
The correctness of the results of this program depend only upon the 
correct coding of the physical equations.  The algebraic machinations 
are repeated until the residual of the fluid differential equations 
and boundary conditions are zero to the requisite order.

{\footnotesize
\verbatimlisting{threed.red}
}

%
%

\bibliographystyle{plain}\bibliography{bib,ajr,new}

\begin{thebibliography}{10}

\bibitem{Atherton73}
R.~W. Atherton and G.~M. Homsy.
\newblock Use of symbolic computation to generate evolution equations and
  asymptotic solutions to elliptic equations.
\newblock {\em J.~Comp. Phys.}, 13:45--58, 1973.

\bibitem{Atherton76}
R.~W. Atherton and G.~M. Homsy.
\newblock On the derivation of evolution equations for interfacial waves.
\newblock {\em Chem. Eng. Comm.}, 2:57--77, 1976.

\bibitem{Batchelor79}
G.~K. Batchelor.
\newblock {\em An introduction to fluid dynamics}.
\newblock Cambridge University Press, 1979.

\bibitem{Bousmar03}
D.~Bousmar and Y.~Zech.
\newblock Large-scale coherent structures in compound channels.
\newblock Technical report, Universiti\'e Catholique de Louvain, 2003.

\bibitem{Bousmar02}
Didier Bousmar.
\newblock {\em Flow modelling in compund channels: momentum transfer between
  main channel and prismatic and non-prismatic floodplains}.
\newblock PhD thesis, Universiti\'e Catholique de Louvain, 2002.

\bibitem{Carr81}
J.~Carr.
\newblock {\em Applications of centre manifold theory}, volume~35 of {\em
  Applied Math. Sci.}
\newblock Springer-Verlag, 1981.

\bibitem{Chang94}
H.~C. Chang.
\newblock Wave evolution on a falling film.
\newblock {\em Annu. Rev. Fluid Mech.}, 26:103--136, 1994.

\bibitem{Chang02}
Hsueh-Chia Chang, Evgenya Demekhin, and Sergey~S. Saprikin.
\newblock Noise-driven wave transitions on a vertically falling film.
\newblock {\em J.~Fluid Mech.}, 462:255--­283, 2002.

\bibitem{Cox93b}
S.~M. Cox and A.~J. Roberts.
\newblock Initial conditions for models of dynamical systems.
\newblock {\em Physica~D}, 85:126--141, 1995.

\bibitem{Decre03}
Michel M.~J. Decr\'e and Jean-Christophe Baret.
\newblock Gravity-driven flows of viscous liquids over two-dimensional
  topographies.
\newblock {\em J. Fluid Mech.}, 487:147­--166, 2003.

\bibitem{Gallay93}
Th. Gallay.
\newblock A center-stable manifold theorem for differential equations in banach
  spaces.
\newblock {\em Commun. Math. Phys}, 152:249--268, 1993.

\bibitem{Grotberg94}
J.~B. Grotberg.
\newblock Pulmonary flow and transport phenomena.
\newblock {\em Annu. Rev. Fluid Mech.}, 26:529--571, 1994.

\bibitem{Gugg63}
H.~W. Guggenheimer.
\newblock {\em Differential geometry}.
\newblock McGraw-Hill Book Company, Inc, 1963.

\bibitem{Haragus95}
M.~H\u{a}r\u{a}gu\c{s}.
\newblock Model equations for water waves in the presence of surface tension.
\newblock {\em Eur. J.~Mech}, 15:471--492, 1996.

\bibitem{Jensen97}
O.~E. Jensen.
\newblock The thin liquid lining of a weakly curved cylindrical tube.
\newblock {\em J.~Fluid Mech}, 331:373--403, 1997.

\bibitem{Kalliadasis94}
S.~Kalliadasis and H.-C. Chang.
\newblock Drop formation during coating of vertical fibres.
\newblock {\em J.~Fluid Mech.}, 261:135--168, 1994.

\bibitem{Kheshgi89}
H.~S. Kheshgi.
\newblock Profile equations for film flows at moderate {R}eynolds numbers.
\newblock {\em AIChE Journal}, 35:1719--1727, 1989.

\bibitem{Kliakhandler01}
I.~L. Kliakhandler, S.~H. Davis, and S.~G. Bankoff.
\newblock Viscous beads on vertical fibre.
\newblock {\em J.~Fluid Mech.}, 429:381--390, 2001.

\bibitem{Lange99}
U.~Lange, K.~Nandakumar, and H.~Raszillier.
\newblock Symbolic computation as a tool for high-order long-wave stability
  analysis of thin film flows with coupled transport processes.
\newblock {\em J.~Comput. Phys.}, 150:1--16, 1999.

\bibitem{Mercer90}
G.~N. Mercer and A.~J. Roberts.
\newblock A centre manifold description of contaminant dispersion in channels
  with varying flow properties.
\newblock {\em SIAM J.~Appl. Math.}, 50:1547--1565, 1990.

\bibitem{Mercer94a}
G.~N. Mercer and A.~J. Roberts.
\newblock A complete model of shear dispersion in pipes.
\newblock {\em Jap. J.~Indust. Appl. Math.}, 11:499--521, 1994.

\bibitem{Miles90a}
J.~Miles and D.~Henderson.
\newblock Parametrically forced surface waves.
\newblock {\em Annu. Rev. Fluid Mech.}, 20:143--165, 1990.

\bibitem{Moriarty91}
J.~A. Moriarty, L.~W. Schwartz, and E.~O. Tuck.
\newblock Unsteady spreading of thin liquid films with small surface tension.
\newblock {\em Phys. Fluids A}, 3(5):733--742, 1991.

\bibitem{Oron97}
A.~Oron, S.~H. Davis, and S.~G. Bankoff.
\newblock Long-scale evolution of thin liquid films.
\newblock {\em Rev. Mod. Phys.}, 69:931--980, 1997.

\bibitem{Perlin00}
Marc Perlin and William~W. Schultz.
\newblock Capillary effects on surface waves.
\newblock {\em Annu. Rev. Fluid Mech.}, 32:241--274, 2000.

\bibitem{Prokopiou91}
T.~Prokopiou, M.~J. Mccready, and H.~C. Chang.
\newblock Wave transitions on horizontal gas sheared liquid films.
\newblock Technical report, Preprint, 1991.

\bibitem{Prokopiou91b}
Th. Prokopiou, M.~Cheng, and H.~C. Chang.
\newblock {Long waves on inclined films at high Reynolds number}.
\newblock {\em J.~Fluid Mech.}, 222:665--691, 1991.

\bibitem{Ribe01}
N.~M. Ribe.
\newblock Bending and stretching of thin viscous sheets.
\newblock {\em J.~Fluid Mech.}, 433:135--160, 2001.

\bibitem{Roberts88a}
A.~J. Roberts.
\newblock The application of centre manifold theory to the evolution of systems
  which vary slowly in space.
\newblock {\em J.~Austral. Math. Soc.~B}, 29:480--500, 1988.

\bibitem{Roberts89}
A.~J. Roberts.
\newblock The utility of an invariant manifold description of the evolution of
  a dynamical system.
\newblock {\em SIAM J.~Math. Anal.}, 20:1447--1458, 1989.

\bibitem{Roberts92}
A.~J. Roberts.
\newblock A sub-centre manifold description of the evolution and interaction of
  nonlinear dispersive waves.
\newblock In L.~Debnath, editor, {\em Nonlinear waves}, chapter~9, pages
  127--156. World Sci, 1992.

\bibitem{Roberts94c}
A.~J. Roberts.
\newblock Low-dimensional models of thin film fluid dynamics.
\newblock {\em Phys. Letts.~A}, 212:63--72, 1996.

\bibitem{Roberts96a}
A.~J. Roberts.
\newblock Low-dimensional modelling of dynamics via computer algebra.
\newblock {\em Computer Phys. Comm.}, 100:215--230, 1997.

\bibitem{Roberts96b}
A.~J. Roberts.
\newblock An accurate model of thin 2d fluid flows with inertia on curved
  surfaces.
\newblock In P.~A. Tyvand, editor, {\em Free-surface flows with viscosity},
  volume~16 of {\em Advances in Fluid Mechanics Series}, chapter~3, pages
  69--88. Comput Mech Pub, 1998.

\bibitem{Roskes69}
G.~J. Roskes.
\newblock Three-dimensional long waves on a liquid film.
\newblock {\em Phys. Fluids}, 13:1440--1445, 1969.

\bibitem{Roy96}
R.~Valery Roy, A.~J. Roberts, and M.~E. Simpson.
\newblock A lubrication model of coating flows over a curved substrate in
  space.
\newblock {\em J.~Fluid Mech.}, 454:235--261, 2002.

\bibitem{Ruschak85}
K.~J. Ruschak.
\newblock Coating flows.
\newblock {\em Annu. Rev. Fluid Mech.}, 17:65--89, 1985.

\bibitem{Schwartz95}
L.~W. Schwartz and D.~E. Weidner.
\newblock Modeling of coating flows on curved surfaces.
\newblock {\em J.~Engrg Maths}, 29:91--103, 1995.

\bibitem{Schwartz95b}
L.~W. Schwartz, D.~E. Weidner, and R.~R. Eley.
\newblock An analysis of the effect of surfactant on the levelling behaviour of
  a thin coating layer.
\newblock {\em Langmuir}, 11:3690--3693, 1995.

\bibitem{Suslov98b}
S.~A. Suslov and A.~J. Roberts.
\newblock Proper initial conditions for the lubrication model of thin film
  fluid flow.
\newblock Technical report, [\url{http://arXiv.org/abs/chao-dyn/9804018}],
  1998.

\bibitem{Watanabe00}
Shinya Watanabe, Vachtang Putkaradze, and Tomas Bohr.
\newblock Integral methods for shallow free-surface flows with separation.
\newblock {\em J.~Fluid Mech.}, 480:233--265, 2003.

\bibitem{Watt94b}
S.~D. Watt and A.~J. Roberts.
\newblock The accurate dynamic modelling of contaminant dispersion in channels.
\newblock {\em SIAM~J. Appl Math}, 55(4):1016--1038, 1995.

\end{thebibliography}

\end{document}